\global\def\draftcontrol{0}

   \def\versionno{real local P2}

\catcode`\@=11

\expandafter\ifx\csname draftcontrol\endcsname\relax\global\def\draftcontrol{0} 
\fi 

{\count255=\time\divide\count255 by 60 
\xdef\hourmin{\number\count255} 
\multiply\count255 by-60\advance\count255 by\time 
\xdef\hourmin{\hourmin:\ifnum\count255<10 0\fi\the\count255}} 
\def\draftdate{\number\month/\number\day/\number\year\ \ \ \hourmin } 

\newcommand\makepapertitle{\par

  \begingroup 
    \renewcommand\thefootnote{\@fnsymbol\c@footnote}%
    \def\@makefnmark{\rlap{\@textsuperscript{\normalfont\@thefnmark}}}%
    \long\def\@makefntext##1{\parindent 1em\noindent 
            \hb@xt@1.8em{%
                \hss\@textsuperscript{\normalfont\@thefnmark}}##1}%
     \newpage 
     \global\@topnum\z@   
     \@makepapertitle 
     \thispagestyle{empty}\@thanks 
  \endgroup 
  \setcounter{footnote}{0}%
  \global\let\thanks\relax 
  \global\let\makepapertitle\relax 
  \global\let\@makepapertitle\relax 
  \global\let\@thanks\@empty 
  \global\let\@author\@empty 
  \global\let\@date\@empty 
  \global\let\@title\@empty 
  \global\let\title\relax 
  \global\let\author\relax 
  \global\let\date\relax 
  \global\let\and\relax 
  \def\version{\let\version\@version\@gobble} 
} 
\def\@makepapertitle{%
  \newpage 
   \ifnum\draftcontrol=1 {} 
   \version\versionno 
   \vskip 5.5em%
   \else 
   \hfill\hbox to 3.5cm {\parbox{5cm}{\@pubnum}\hss}%
   \vskip 6.5em%
   \fi 
   \begin{center}%
   \let \footnote \thanks 
      {\hskip -0\textwidth \hbox to 1\textwidth%
        {\centerline{\Large\bf{\noindent\@title}}}}%
     \vskip 2em%
     {\normalsize
       \lineskip .5em%
       \begin{tabular}[t]{c}%
         \@author 
       \end{tabular}\par}%
     \vskip 1.5em%
     {\@bstract}%
     \end{center}%
     \vfill
     \@date%
     \vskip 1.5em%
   \par 
} 

\gdef\@pubnum{} 
\def\pubnum#1{%
  \gdef\@pubnum{#1}} 

\gdef\@bstract{} 
\def\Abstract#1{%
  \gdef\@bstract{%
   \parbox{\textwidth-0pc}{%
   \centerline{\bf Abstract}\penalty1000 
   \noindent
   \renewcommand\baselinestretch{1.0} 
   {#1}}} 
} 

\gdef\@email{}
\def\email#1{%
   \gdef\@email{%
   Email: {\tt #1}}
}

\def\ps@paper{\let\@mkboth\@gobbletwo%
     \ifnum\draftcontrol=1 
        \def\@oddfoot{\hbox to \textwidth{\tiny \versionno \hfil\tiny\draftdate}%
        \hskip -\textwidth \hbox to \textwidth{\hfil\rm\thepage\hfil}}%
     \else\def\@oddfoot{\hbox to \textwidth{\hfil\rm\thepage\hfil}} 
     \fi 
     \let\@evenfoot\@oddfoot 
} 

\def\body{\clearpage 
          \pagestyle{paper} 
        } 
\newenvironment{acknowledgments}{%
\vskip 3.25ex 
\addcontentsline{toc}{section}{Acknowledgments}
\noindent {\bf Acknowledgments} 
} 

\def\@version#1{\ifnum\draftcontrol=1 
\typeout{}\typeout{#1}\typeout{} 
\vskip3mm\centerline{\hbox{\fbox{\normalsize{\tt DRAFT -- #1 -- } 
                   {\draftdate}}}}\vskip3mm 
\fi} 
\let\version\@version 
\long\def\eqlabel#1{\ifnum\draftcontrol=1 
                    \tag@false  
                    \tag*{(\theequation) \hbox to -0.2cm{\hspace{0cm}\small{#1}\hss}} 
                    \refstepcounter{equation}  
                    \edef\@currentlabel{\theequation} 
                    \ltx@label{#1}          
                    \else 
                    \label{#1} 
                    \fi 
                    } 
\let\st@bibitem\@bibitem 
\let\st@lbibitem\@lbibitem 
\ifnum\draftcontrol=1 
  \def\@bibitem#1{%
    \st@bibitem{#1}\a@@label{#1}\ignorespaces} 
  \def\@lbibitem[#1]#2{%
    \st@lbibitem[#1]{#2}\a@@label{#2}\ignorespaces} 
  \def\a@@label#1{%
    \gdef\a@lab{\smash{\normalfont\small#1}} 
    \ifvmode 
      \if@inlabel 
        \global\setbox\@labels\hbox{%
          \llap{\a@lab\let\a@lab\relax 
                \kern\@totalleftmargin\kern\marginparsep}%
          \box\@labels}%
      \fi 
    \fi} 
\fi 

\documentclass[12pt,letterpaper]{article} 

\usepackage{amsmath,bm,amsfonts,amssymb,array,calc,amsthm,rotating}
\usepackage{epsfig,psfrag} 
\usepackage{rotating}
\usepackage{amscd}
\usepackage{graphicx}
\usepackage{color}
\usepackage[colorlinks=false]{hyperref}

\renewcommand\baselinestretch{1.25} 
\renewcommand\section{\@startsection {section}{1}{\z@}%
                                   {-3.5ex \@plus -1ex \@minus -.2ex}%
                                   {2.3ex \@plus.2ex}%
                                   {\normalfont\large\bfseries}} 
\renewcommand\subsection{\@startsection{subsection}{2}{\z@}%
                                   {-3.25ex\@plus -1ex \@minus -.2ex}%
                                   {1.5ex \@plus .2ex}%
                                   {\normalfont\normalsize\bfseries}} 
\renewcommand\subsubsection{\@startsection{subsubsection}{3}{\z@}%
                                   {-3.25ex\@plus -1ex \@minus -.2ex}%
                                   {1.5ex \@plus .2ex}%
                                   {\normalfont\normalsize\it}} 
\renewcommand\paragraph{\@startsection{paragraph}{4}{\z@}%
                                   {-3.25ex\@plus -1ex \@minus -.2ex}%
                                   {1.5ex \@plus .2ex}%
                                   {\normalfont\normalsize\bf}} 
\renewcommand\subparagraph{\@startsection{subparagraph}{5}{\z@}%
                                   {-1.25ex\@plus -1ex \@minus -.2ex}%
                                   {0ex \@plus .2ex}%
                                   {\normalfont\normalsize\it}}

\numberwithin{equation}{section}

\long\def\@makecaption#1#2{%
  \vskip\abovecaptionskip
  \sbox\@tempboxa{{\bf #1:} #2}%
  \ifdim \wd\@tempboxa >\hsize
    {\small\bf #1:} {\small #2}\par
  \else
    \global \@minipagefalse
    \hb@xt@\hsize{\hfil\box\@tempboxa\hfil}%
  \fi
  \vskip\belowcaptionskip}

\renewcommand*\l@section[2]{%
  \ifnum \c@tocdepth >\z@
    \addpenalty\@secpenalty
    \addvspace{.5em \@plus\p@}%
    \setlength\@tempdima{1.5em}%
    \begingroup
      \parindent \z@ \rightskip \@pnumwidth
      \parfillskip -\@pnumwidth
      \leavevmode \bfseries
      \advance\leftskip\@tempdima
      \hskip -\leftskip
      #1\nobreak\hfil \nobreak\hb@xt@\@pnumwidth{\hss #2}\par
    \endgroup
  \fi}
\renewcommand*\l@subsection{\addvspace{.0em \@plus\p@}\@dottedtocline{2}{1.5em}{2.3em}}
\renewcommand*\l@subsubsection{\addvspace{-.2em \@plus\p@}\@dottedtocline{3}{3.8em}{3.2em}}

\def\hepth#1{\href{http://xxx.arxiv.org/abs/hep-th/#1}{{arXiv:hep-th/#1}}}

\def\mathsg#1{\href{http://xxx.arxiv.org/abs/math.SG/#1}{{arXiv:math.sg/#1}}}
\def\mathag#1{\href{http://xxx.arxiv.org/abs/math.AG/#1}{{arXiv:math.ag/#1}}}

\def\arxiv#1#2{\href{http://xxx.arxiv.org/abs/#1}{{arXiv:#1 [#2]}}}

\definecolor{refcol}{rgb}{0.2,0.2,0.8}
\definecolor{eqcol}{rgb}{.6,0,0}
\definecolor{purple}{cmyk}{0,1,0,0}

\gdef\@citecolor{refcol}
\gdef\@linkcolor{eqcol}
\def\colorlinkspurple{\gdef\@urlcolor{purple}}
\def\colorlinksblue{\gdef\@urlcolor{blue}}
\def\colorlinksred{\gdef\@urlcolor{red}}

\def\ie{{\it i.e.}} 
\def\eg{{\it e.g.}}

\def\revise#1       {\raisebox{-0em}{\rule{3pt}{1em}}%
                     \marginpar{\raisebox{.5em}{\vrule width3pt\ 
                     \vrule width0pt height 0pt depth0.5em 
                     \hbox to 0cm{\hspace{0cm}{%
                     \parbox[t]{4em}{\raggedright\footnotesize{#1}}}\hss}}}}

\def\calo         {{\cal O}}

\def\projective   {{\mathbb P}} 
 
\def\reals        {{\mathbb R}} 
\def\zet          {{\mathbb Z}} 

\def\RP{\reals\projective}

\def\del          {\partial} 
 
\def\ee           {{\it e}} 
\def\ii           {{\it i}}

\newcommand\topa[2]{\genfrac{}{}{0pt}{2}{\scriptstyle #1}{\scriptstyle #2}}

\def\sqr#1#2{{\vcenter{\vbox{\hrule height.#2pt   
 \hbox{\vrule width.#2pt height#1pt \kern#1pt 
 \vrule width.#2pt}\hrule height.#2pt}}}}

\newcommand{\G}{\Gamma}
\newcommand{\h}{\hat}
\newcommand{\A}{\mathbb A}

\newcommand{\beq}{\begin{equation}}
\newcommand{\eq}{\end{equation}}
\newcommand{\Gx}{\mathcal G^{(\chi)}}
\newcommand{\Ms}{\overline{\mathcal M}}
\renewcommand{\t}{\tilde}
\newcommand{\F}{\mathcal F}
\newcommand{\R}{\mathcal R}
\newcommand{\K}{\mathcal K}
\renewcommand{\P}{\mathbb P}
\newcommand{\Z}{\mathbb Z}
\newcommand{\E}{\mathcal E}
\renewcommand{\O}{\mathcal O}

\newcommand{\T}{\mathbb T}
\newcommand{\C}{\mathbb C}
\newcommand{\N}{\mathcal N}
\renewcommand{\o}{\omega}
\renewcommand{\d}{\partial}
\renewcommand{\b}{\bar}
\newcommand{\D}{\Delta}
\renewcommand{\i}{i}
\renewcommand{\th}{\theta}

\catcode`\@=12 

\begin{document} 


\title{The Real Topological String on a local Calabi-Yau}

\pubnum{
\arxiv{0902.0616}{hep-th}\\
CERN-PH-TH-2009-022 \\
LMU-ASC 06/09
}
\date{February 2009}

\author{
Daniel Krefl$^{a,b}$ and Johannes Walcher$^{b}$ \\[0.2cm]
\it ${}^{a}$ 
\href{http://www.theorie.physik.uni-muenchen.de/index.html}
{Arnold Sommerfeld Center for Theoretical Physics}, LMU Munich, Germany \\
\it $^{b}$ 
\href{http://ph-dep-th.web.cern.ch/ph-dep-th/}{PH-TH Division}, CERN, Geneva, Switzerland
}

\Abstract{
We study the topological string on local $\P^2$ with O-plane and D-brane at its 
real locus, using three complementary techniques. 
In the A-model, we refine localization on the moduli space of maps with
respect to the torus action preserved by the anti-holomorphic involution. This
leads to a computation of open and unoriented Gromov-Witten invariants that can
be applied to any toric Calabi-Yau with involution. We then show that the full 
topological string amplitudes can be reproduced within the topological vertex
formalism. We obtain the real topological vertex with trivial fixed leg. Finally,
we verify that the same results derive in the B-model from the extended holomorphic 
anomaly equation, together with appropriate boundary conditions. The expansion at the conifold
exhibits a gap structure that belongs to a so far unidentified universality class.
}

\makepapertitle

\body

\version\versionno

\vskip 1em

\tableofcontents

\section{Introduction and Overview}

There has been a lot of progress in closed and open topological string theory 
in the last couple of years. The improved understanding concerns in particular local 
(non-compact) backgrounds defined by toric Calabi-Yau manifolds together with toric 
branes on top. While many lessons were learned (for reviews see for instance 
\cite{Neitzke:2004ni,Marino:2004uf}), it has long not been clear how they would 
apply to compact backgrounds, which indeed remain the challenging case to 
understand in general. 

Recently, it has become clearer that there are significant qualitative distinctions 
between the non-compact and compact settings. Perhaps the most dramatic additional
ingredient is a topological analogue \cite{Walcher:2007qp} of the tadpole cancellation 
condition familiar from the type II superstring. In particular, a satisfactory 
BPS interpretation of the topological string amplitudes requires that one consider 
topological string orientifolds, whose charge precisely cancels that of the 
background D-branes. We will be considering O-planes and D-branes defined via 
the fixed locus of an anti-holomorphic involution, and will refer to the resulting 
theory as the real topological string.

Issues such as tadpole cancellation might seem to cast doubt on the general 
applicability of any local lessons. As an example, large-$N$ dualities cannot 
be useful if the total D-brane charge is restricted. In the present paper, we 
show that the situation is actually slightly better. Specifically, we will study
the real topological string on the local Calabi-Yau manifold given by the
canonical bundle over the projective plane (local $\P^2$). Among our main findings
are several parallels both with the usual toric story, as well as 
with the real topological string on a compact manifold. We hope that these connections 
will prove useful for both lines of investigation.

A physical motivation for the importance of the real topological string comes from
considering the combined open and closed type IIA superstring with orientifold 
projection, which is a well-known playground for string phenomenology. 
Recall that this orientifold projection is the gauging of a discrete symmetry $I\circ P$, 
where $I$ is an anti-holomorphic involution of the internal background $X$ and $P$ 
denotes parity reversal on the string world-sheet. The world-sheets of the orientifolded 
theory then have general topology, in the sense that they can be oriented or unoriented
and may possess boundaries and/or cross-caps. As is well known, one can represent these
world-sheets as quotients $\h\Sigma/\sigma$ of a closed oriented world-sheet $\h\Sigma$ 
by an anti-holomorphic involution $\sigma$. The equivalence class of $\sigma$
determines the topology of $\h\Sigma/\sigma$. In the non-perturbative (in $\alpha'$) 
sector of such orientifolded type IIA theories, one has to consider world-sheet 
instantons with general topology, \ie, maps from Riemann surfaces with or without 
boundaries and cross-caps into target-space equipped with involution. As usual, 
the summation of world-sheet instantons is best done by considering the 
topological theory, which is the interest of the present paper. 

We begin by recalling the main features of the setup of \cite{Walcher:2007qp}, and 
fix some notation, before summarizing our main results.

The target space that we shall study is the local Calabi-Yau $X=\calo_{\P^2}(-3)$. The
involution $I$ defining the orientifold projection is simply complex conjugation. The 
fixed point locus, $L$, on which we shall wrap one D-brane, is the real version of the 
canonical bundle, and can be thought of as the real line bundle defined by the orientation 
bundle over $\RP^2$. (Note that, as special Lagrangian, $L$, itself, is oriented.)

The central object to compute is the total, or combined open-closed-unoriented\footnote{To 
emphasize one point again: When the tadpole cancelling D-branes are put right on top 
of the orientifold plane, we refer to the theory as ``real''. Certain of the present 
definitions are good somewhat more generally.} topological string free energy, 
which in a perturbative expansion can be written as:
\beq\eqlabel{inteq0a}
\mathcal G=\sum_{\chi=-2}^\infty \Gx \lambda^{\chi},
\eq
Here $\Gx$ is the contribution at order $\chi$, and $\lambda$ is the string
coupling. In general, the ${\mathcal G}$ and 
$\Gx$ depend on closed and open string moduli, which in the A-model consist of
K\"ahler moduli of $X$ and complexified Wilson lines on the D-branes. In the example 
of interest, we have $H_2(X;\zet)=\zet$, and $H_1(L;\zet)=\zet_2$, so we have 
one continuous closed string modulus, denoted by $t\equiv\log q$, and one discrete 
open string modulus, $\epsilon=\pm 1$. Thus,
\begin{equation}
\Gx = \Gx(t,\epsilon)\,.
\end{equation}
On general grounds, one expects to be able to compute $\Gx$ by summing contributions
from individual world-sheet topologies, \footnote{In contrast to the physically 
motivated normalization of $\Gx$ used in \cite{Walcher:2007qp}, we chose here a 
different normalization which is more convenient for practical computations.}
\beq
\eqlabel{inteq2}
\Gx=\sum_{2g+h-2=\chi}\F^{(g,h)}+\sum_{2g+h-1=\chi}\R^{(g,h)}+\sum_{2g+h-2=\chi}\K^{(g,h)}~.
\eq
Namely, $\F^{(g,h)}$ (with $\F^{(g)}\equiv\F^{(g,0)}$) is the contribution of oriented 
genus $g$ surfaces with $h$ boundaries, $\R^{(g,h)}$ is the contribution of unoriented 
genus $g$ surfaces with $h$ boundaries and an odd number of cross-caps (note that one 
can trade three cross-caps for a handle plus a cross-cap) and $\K^{(g,h)}$ comes from 
unoriented genus $g$ surfaces with $h$ boundaries and an even number of cross-caps.
(Note that one can trade two cross-caps for a Klein handle, that is a handle with 
orientation reversal. The genus $g$ in $\K^{(g,h)}$ refers to the number 
of handles plus the number of Klein handles, with at least one Klein handle.)

Moreover, each of those contributions in \eqref{inteq2} should be computable by 
counting the number of maps from the appropriate surfaces into the background, 
similar to the expansion of the closed string free energy
\beq
\F=\sum_{g=0}^\infty \F^{(g)}\lambda^{2g-2}~,
\eq
with (ignoring constant map contributions polynomial in $t=\log q$)
\begin{equation}
\F^{(g)}= \sum_{d} \tilde n^{(g)}_d q^d~,
\end{equation}
where the sum is over (positive) $d\in H_2(X;\zet)$, and $\tilde n^{(g)}_d$
are the rational Gromov-Witten invariants. For future reference, we 
note the following expansion of $\F$ in terms of integer BPS degeneracies,~\ie, 
Gopakumar-Vafa invariants $N_d^{(g)}$,
\beq
\eqlabel{standardgv}
\F=\sum_{g,d,k}N_d^{(g)}\frac{1}{k}\left(2\sinh\frac{\lambda k}{2}\right)^{2g-2} q^{kd}~.
\eq

In hindsight (say if one is given the answer by some other means) it is not
necessarily clear how to disentangle the individual contributions in \eqref{inteq2}.
Cancellation of the O-plane tadpole allows wrapping only a single D-brane on $L$,
so we only have one discrete open string modulus $\epsilon$ at our disposal. This
only allows distinguishing whether $h$ is even or odd. In some of our computations,
however, there are ways to effectively introduce arbitrary numbers of brane-antibrane
pairs, each with their discrete Wilson line degree of freedom. This allows keeping track
of individual world-sheet topologies. Then we may write for $h>0$:
\beq
\eqlabel{disentangle}
\begin{split}
\F^{(g,h)}&=\sum_{d\equiv h\bmod 2}\t n^{(g,h)}_d q^{d/2} \epsilon^h\,,\\
\K^{(g,h)}&=\sum_{d\equiv h\bmod 2}\t n^{(g,h)_k}_d q^{d/2} \epsilon^h\,,
\end{split}
\eq
where the $\t n^{(g,h)}_d$ and $\t n^{(g,h)_k}_d$ are appropriate open and unoriented
Gromov-Witten invariants. In these expressions, $d$ refers to the relative homology 
class in $H_2(X,L)$, or in the case of unoriented surfaces, the homology class 
of the covering map.

More precisely, to write \eqref{disentangle}, one has to assume a certain prescription
to deal with homologically trivial boundaries, which we will recall below. This
prescription, together with the map $H_2(X,L)\to H_1(L)$ also explains the 
restriction to $d\equiv h\bmod 2$, and entails the vanishing of the $\R^{(g,h)}$
in our model.

Independently of such assumptions, we can isolate the contribution from purely oriented
closed strings (because that is known from before the orientifold projection!). Thus
we define the amplitude ${{\mathcal G}'}^{(\chi)}$
\beq
{{\mathcal G}'}^{(\chi)}=\Gx-\left\{
\begin{matrix}
\F^{(\frac{\chi}{2}+1,0)}&\text{for}~\chi~\text{even}
\\
0&\text{for}~\chi~\text{odd}
\end{matrix}
\right.~,
\eq
which can be seen to have an expansion of the form
\beq\label{inteq2a}
{{\mathcal G}'}^{(\chi)}=\sum_{d\equiv \chi\bmod 2}n'^{(\chi)}_{d} q^{d/2}\epsilon^\chi\,,
\eq
in terms of rational numbers ${n'}^{(\chi)}_d$, which one might call real Gromov-Witten 
invariants. As found in \cite{Walcher:2007qp}, the combined open-closed-unoriented 
topological string free energy without oriented closed string contribution,
\beq
{{\mathcal G}'}=\sum_\chi{{\mathcal G}'}^{(\chi)}\lambda^\chi~,
\eq
possesses an expansion with integer coefficients $N'^{(\chi)}_d$, similar to
that of the $\F^{(g)}$ in eq.\ \eqref{standardgv}
\beq
\eqlabel{possesses}
\frac{1}{2}{{\mathcal G}'}=\sum_{\substack{d\equiv \chi~\text{mod}~2\\ k~\text{odd}}}
N'^{(\chi)}_d 
\frac{1}{k}\left(2\sinh\frac{\lambda k}{2}\right)^\chi q^{kd/2}~ \epsilon^\chi~.
\eq
The $N'^{(\chi)}_d$ should be seen as a real version of Gopakumar-Vafa invariants,
counting real degree $d$ curves. Physically, they also count dimensions of Hilbert spaces of
appropriate BPS objects \cite{oova}.

A nice property of the real topological string is that local and compact 
backgrounds are more closely related (the real brane is usually non-toric 
in local settings), and hence one can learn more for the compact case from the local 
real case than from the usual toric open topological string. On the other hand, some 
calculation techniques from the local toric case remain applicable, as we will explain 
presently.

For the model at hand, the individual contributions in \eqref{disentangle} can be 
explicitly calculated via localization on the moduli space of stable maps, as 
performed by Kontsevich to calculate $\t n_d^{(0)}$ \cite{Kontsevich:1994na}, 
generalized by several authors to the open string case 
\cite{Graber:2001dw,Diaconescu:2003dq} and recently completed by the inclusion of 
unoriented strings \cite{Walcher:2007qp}. Especially, localization was used to compute 
various oriented amplitudes for our model of interest,~\ie, local $\P^2$, in 
\cite{Klemm:1999gm,Mayr:2002zi}. The essential point that allows the extension 
to the real case, in this and other models, is that although the real brane is 
usually non-toric, it is often left invariant by the action of at least a 
one-dimensional torus. This is enough for localization to apply. (A toric 
brane in the usual sense is by definition always invariant under a two-dimensional 
torus.) We will review and apply this approach in section \ref{loc} to calculate 
the individual contributions to the topological amplitudes of local $\P^2$ for 
some higher $\chi$ and $d$. 

In section \ref{realvertex}, we will take a different approach to the same problem 
and derive the total topological string amplitudes via a real version of the 
topological vertex. Recall that the standard topological vertex solves the closed 
topological string (with background toric branes) on local toric Calabi-Yau threefolds 
by evaluating a certain cubic field theory on the toric diagram of the Calabi-Yau 
viewed as a Feynman diagram \cite{Aganagic:2003db}. Applications of the topological
vertex to orientifolds have been considered before, such as in 
\cite{Bouchard:2004iu,Bouchard:2004ri}. In these works, the
involution defining the orientifold was taken to be freely acting. The main new 
feature in our study is that we deal with a non-empty orientifold plane. This 
also requires the introduction of a specific D-brane into the background on top of 
the O-plane. An orientifold model that can be solved with these techniques of either 
localization or the topological vertex has the property that the toric diagram has
an involutive symmetry to define the orientifold projection. (Toric Calabi-Yaus, which
are rigid, are always invariant under complex conjugation, but unless this can be
dressed with a symmetry of the toric diagram, no toric symmetry will be preserved.)
There are then several possible cases for the fixed point locus. A new feature arises
when there are vertices fixed under the involution of the toric diagram, and one 
then has to distinguish whether the fixed leg (of which there is necessarily exactly 
one) ending on the fixed vertex is ``external'' to the toric diagram or not. We 
will call the requisite transition amplitude the ``real topological vertex''. By 
studying real local $\P^2$, we will be able to deduce the real vertex in which 
the fixed leg is external. Since our main aim here is a proof of principle, we 
will not try to go beyond that. It is conceivable that a more complete theory exists.

Both localization and the topological vertex fail in general for compact models. The 
only tool available which works also in the compact setup, is mirror symmetry 
together with the (extended) holomorphic anomaly equations of \cite{Bershadsky:1993cx,
Walcher:2007tp}. This approach has the notorious problem that one has to fix the 
holomorphic ambiguity (boundary conditions on moduli space) at each order in 
perturbation theory. In the closed topological string it has been shown 
\cite{Huang:2006si,Huang:2006hq}  that detailed information about the singularity 
structure at the conifold locus can be carried over to compact models and leads to 
a very efficient solution scheme up to very high genus. For non-compact models,
the same structure leads to complete integrability (for an explicit example, see 
\cite{Haghighat:2008gw}). It is natural to look for a similar structure also in 
the real topological string, and indeed we will make a find, see section \ref{Bmodel}.

Mirror symmetry and the holomorphic anomaly have the advantage that they give an
answer to all orders in the instanton expansion, but the disadvantage that they
are limited to an order-by-order calculation in the string coupling expansion. On
the other hand, the topological vertex gives an all-order result in the string coupling,
but in practical computations is limited to the first few orders in the instanton
expansion. Finally, localization is an order-by-order computation in both 
directions, and also computationally rather challenging. What it has going for it
is that of the three techniques we study, it is the one that is likely easiest to
put on a rigorous mathematical foundation.

Some more concluding words with a sketch of possible directions of follow-up research are 
offered in section \ref{Conc}. Finally, the results for the real Gopakumar-Vafa invariants 
$N'^{(\chi)}_d$ of local $\P^2$  are collected in appendix \ref{secA}.

\section{The A-model}
\label{loc}

In this section, we explain the computation of open and unoriented Gromov-Witten
invariants of the real topological string on local $\P^2$ using localization on 
the space of maps. For the reader's convenience, we firstly recall some basics about 
the localization calculation for pure closed string world-sheets. A more detailed 
exposition can be found in standard textbooks on mirror symmetry or in the original 
works \cite{Kontsevich:1994na,Graber:1997}. We then discuss the extension to open 
and unoriented world-sheets developed in \cite{Walcher:2007qp}. Especially, we will 
work out in more detail some technical issues which are important at higher degree 
and genus. Some actual results of our calculations are listed in appendix \ref{secA}. 
The reader not interested in explicit A-model computations may safely skip this section.

\subsection{Localization}

We first briefly recall the basics of how to calculate the pure oriented closed string 
contribution $\t n_d^{(g,0)}$ for local $\P^2$ via localization.

Define $\Ms^{\h\Sigma}_d\equiv \Ms_{\h{g},0}(d,\P^2)$ as the moduli space of stable maps 
$\h f:\h\Sigma\rightarrow \P^2$ from genus $\h g$ curves into $\P^2$ with image of 
degree $d\in H_2(\P^2,\Z)$. Let $\mathbf{e}(\E_d)$ be the Euler class of the bundle 
$\E_d=H^1(\Sigma^g,f^* \O(-3))$ over $\Ms^{\h\Sigma}_d$. Then, the Gromov-Witten 
invariants $\t n^{\h\Sigma}_d \equiv \t n_d^{(\h g,0)}$ are given by 
\beq
\eqlabel{loceq1}
\t n^{\h\Sigma}_d=\int_{\Ms^{\h\Sigma}_d} \mathbf {e}(\E_d).
\eq
These integrals can be evaluated by the Atiyah-Bott localization formula.

To this end, consider the $\h\T=(\C^*)^3$ group action on $\P^2$. The fixed points of 
$\h\T$ on $\P^2$ are the three points $p_i$ given by the projectivization of the $i$-th 
coordinate line of $\C^{3}$. The only curves invariant under $\h\T$ are the three lines 
$l_{ij}$ joining the $p_i$. The $\h\T$ action can be pulled back to an action on 
$\Ms^{\h\Sigma}_d$. We will denote the $\h\T$-invariant subspace of $\Ms^{\h\Sigma}_d$ 
as $^{\h\T}\Ms^{\h\Sigma}_d$. Since a point in $^{\h\T}\Ms^{\h\Sigma}_d$ is a map of a 
genus $\h g$ curve $\h \Sigma$ to the $\h\T$-invariant locus in $\P^2$, we immediately 
deduce that $\h \Sigma$ can only consist of the union of a certain number of $n_v$-pointed 
irreducible genus $g_v$ curves $C^{(g_v)}_{v,n_v}$ joined together by $2$-pointed
spheres. The $C^{(g_v)}_{v,n_v}$ are contracted to one of the three points $p_i$,  
while the spheres are mapped to the $l_{ij}$.
It follows that each map $\h f$ can be represented 
combinatorially as a connected graph, \ie, to each map $\h f$ we associate a 
graph $\h\G$ by identifying each contracted component of $\h\Sigma$ with a 
decorated vertex, where the decoration is given by the genus of the component 
and the point $p_i$ it maps to in target space. The spheres joining the contracted 
components are then identified with edges joining the corresponding vertices, 
where each edge is decorated with the degree (\ie, the multi-cover) of the map 
which sends the corresponding sphere to  $l_{ij}$.

Thus, we have a map which associates to each point  in $^{\h\T}\Ms^{\h\Sigma}_d$ a 
decorated graph. Note that the map is not one-to-one, but rather each graph $\h\G$ 
corresponds to a subspace $\h{M}_d^{\h\G}\subset~ ^{\h\T}\Ms^{\h\Sigma}_d$.

In order to see this, observe that each vertex of the graph $\h\G$ comes with the moduli 
space of an $n_v$-pointed genus $g_v$ curve, usually denoted as $\Ms_{g_v,n_v}$. Hence, 
each graph corresponds to the moduli space $\Ms_{\h\G}$ given by
\beq
\Ms_{\h\G}=\prod_v \Ms_{g_v,val(v)}~.
\eq
Obviously, there exists a map $\gamma_{\h\G}:\Ms_{\h\G}\rightarrow \h{M}_d^{\h\G}$, 
which is however not an isomorphism. In order to obtain an isomorphism, we need to quotient 
by the automorphism group of $\Ms_{\h\G}$ given by $\A_{\h\G}={\rm Aut}(\h\G)\ltimes\prod_e
\Z_{d_e}$, where ${\rm Aut}(\h\G)$ is the automorphism group of $\h\G$ as a decorated 
graph. 

Thus, we have
\beq
^{\h\T}\Ms^{\h\Sigma}_d\cong\bigcup_{\h\G}\left(\Ms_{\h\G}/{\A_{\h\G}}\right)~,
\eq
where the union is over the set of all non-isomorphic graphs $\h\G$ whose topology and 
decoration fulfill the following criteria:
\begin{itemize}

\setlength{\itemsep}{0cm} 

\item $\sum_e d_e=d$.
\item $1-|v|+|e|+\sum_v g_v=\h g$, where $|v|$ and $|e|$ is the number of vertices and 
edges, respectively.
\item $i(v_a)\neq i(v_b)$, for $v_a$ connected to $v_b$, where $i(v_j)$ encodes the 
point in target space the corresponding component maps to.
\end{itemize}

Applying the Atiyah-Bott localization formula then tells us that we can evaluate 
\eqref{loceq1} via a sum over graphs:
\beq
\eqlabel{loceq2}
\t n^{\h\Sigma}_d=\sum_{\h\G}\frac{1}{|\A_{\h\G}|}\int_{\Ms_{\h\G}}
\frac{\mathbf{e}(i^*\E_d)}{\mathbf{e}(\N_{\h\G}^{vir})}~,
\eq
where $|\A_{\h\G}|$ is the order of the group $\A_{\h\G}$.

Explicit expressions for $\mathbf{e}(i^*\E_d)$ and $\mathbf{e}(\N_{\h\G}^{vir})$ 
in equivariant cohomology have been derived in \cite{Graber:1997}. We restate them 
here for convenience:
\beq
\eqlabel{loceq3}
\mathbf{e}(i^*\E_d)=\prod_v \lambda_{i(v)}^{val(v)-1}P_{g(v)}(\Lambda_{i(v)})
\prod_e\prod_{m=1}^{3d_e-1}\left[\Lambda_{i(e)}+\frac{m}{d_e}(\lambda_{i(e)}-
\lambda_{j(e)})\right]~,
\eq
\beq
\eqlabel{loceq4}
\begin{split}
\frac{1}{\mathbf{e}(\N_\G^{vir})}=&\prod_e \frac{(-1)^{d_e}d_e^{2d_e}}{(d_e!)^2
(\lambda_{i(e)}-\lambda_{j(e)})^{2d_e}}\prod^{d_e}_{\substack{k\neq i(e),j(e)\\a=0}}
\frac{1}{\frac{a}{d_e}\lambda_{i(e)}+\frac{d_e-a}{d_e}\lambda_{j(e)}-\lambda_k}\\
&\times\prod_v\prod_{j\neq i(v)}(\lambda_{i(v)}-\lambda_j)^{val(v)-1}\\
&\times\left\{\begin{matrix}
\prod_v\left[\left(\sum_Fw_F^{-1}\right)^{val(v)-3}\prod_{F\ni v}w_F^{-1}\right]& 
\text{for $g(v)=0$}\\
\prod_v\prod_{j\neq i(v)}P_{g(v)}(\lambda_{i(v)}-\lambda_j)\prod_{F\ni v}\frac{1}
{w_F-\kappa_F}& \text{for $g(v)\geq 1$}
\end{matrix}
\right.~,
\end{split}
\eq
with
\beq
\begin{split}
w_F&= (\lambda_{i(F)}-\lambda_{j(F)})/d_e~,\\
\Lambda_i&=\lambda_1+\lambda_2+\lambda_3-3\lambda_i~,\\
P_g(x)&=\sum_{r=0}^g c_{g-r}(E^*) x^r~,
\end{split}
\eq
where $i(e)$ and $j(e)$ refer to the target space points the vertices attached to the 
edge $e$ map to, $F$ runs over the set of  flags of a vertex, that is, all pairs $(v,e)$ 
for a fixed vertex $v$ with $e$ ending on $v$. For a flag, we have $i(F)=i(v)$ and
$j(F)$ refers to the other end point of $e$. Finally, $E$ is the Hodge bundle, 
$\kappa_F$ is a gravitational descendant and $\lambda_i$ are the torus weights.

Thus, the integration in equation (\ref{loceq2}) boils down to the evaluation of Hodge 
integrals, for which one can use Faber's algorithm \cite{Faber:1997}.

\subsection{Orientifolded localization}
\label{oloc}

In order to calculate the remaining contributions to $\Gx$ via localization, 
one would like to replace $\Ms^{\h\Sigma}_d$ by something like the moduli space 
$\Ms^{\Sigma}_d$ of stable maps $f:\Sigma\rightarrow C$ from curves $\Sigma$ of 
Euler characteristic $\chi$ (with boundaries and cross-caps) into $\P^2$ with 
image $d$ in the relative homology group $d\in H_2(\P^2,L;\Z)$. 

The proper mathematical definitions related to $\Ms^{\Sigma}_d$ have so far not been 
given, except when $\Sigma$ is the disk \cite{PSW}. Nevertheless, and following
\cite{Walcher:2007qp}, we can give a computational scheme that allows the
evaluation of a putative virtual fundamental class of $\Ms^{\Sigma}_d$, after
localization. The main reason for this simplification is that after implementing the 
tadpole cancellation condition of \cite{Walcher:2007qp}, we effectively 
only need to count maps that send any boundary to a non-trivial one-cycle on $L$,
and that do not contract any cross-caps. As a result, we have to deal only with 
moduli spaces of $n$-pointed genus $g$ curves, as without 
orientifold projection, and also avoid potentially dangerous regions in moduli 
space where a node lies right on top of the orientifold-plane.

\begin{figure}
\psfrag{1}[cc][][1]{a)}
\psfrag{2}[cc][][1]{b)}
\psfrag{3}[cc][][1]{c)}
\psfrag{A}[cc][][0.75]{$\P^2$}
\psfrag{a}[cc][][0.5]{$p_3$}
\psfrag{b}[cc][][0.5]{$p_2$}
\psfrag{c}[cc][][0.5]{$p_1$}
\begin{center}
\includegraphics[scale=0.3]{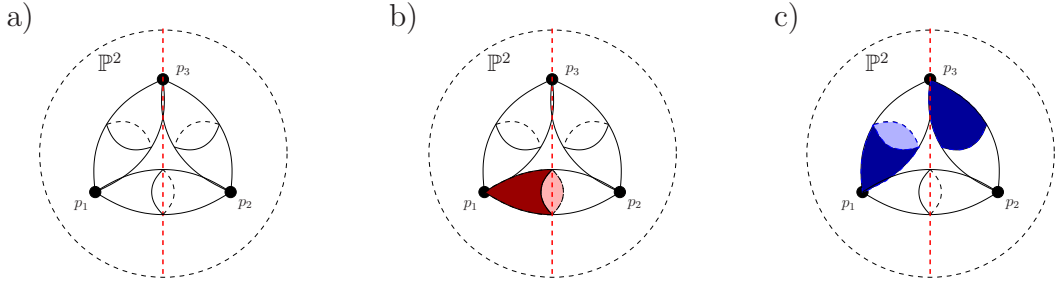}
\caption{a) The orientifold is chosen to act on $\P^2$ such that the $\h \T$ fixed 
points $p_1$ and $p_2$ are identified, while $p_3$ is mapped to itself. The sketched 
(football-shaped) spheres correspond to the lines $l_{ij}$. b) The line $l_{12}$ 
can be mapped to from either a disk or a cross-cap. c) The line $l_{13}$ corresponds in the 
quotient either to a 2-sphere by gluing disks of different color or to a Klein handle 
by gluing two disks of the same color.}
\label{olocfig1}
\end{center}
\end{figure}

To begin, we choose the involution $I$ such that it is maximally compatible with the 
covering space action $\h \T$. This means that the projection leaves a subtorus 
$\T\cong\C^*\subset\h \T$ intact. Such an $I$ identifies two of the three covering 
space fixed-points $p_i$ and as well two of the fixed-lines $l_{ij}$. We arrange 
it such that $p_1$ is identified with $p_2$. The corresponding action $I$ is 
sketched in figure \ref{olocfig1}a. 
We infer that $l_{12}$ is mapped to itself and can receive a disk or a cross-cap, 
as sketched in figure \ref{olocfig1}b. Note that one can glue two of these disks or two 
cross-caps to obtain a 2-sphere or a Klein handle, respectively. The line $l_{13}$ 
can correspond to either a 2-sphere or a Klein handle. How that Klein handle occurs 
is sketched in figure \ref{olocfig1}c. In detail, one half of the line can be thought
to correspond to the line $l_{13}$ while the other half comes from the mirror line 
$l_{23}$. 

As in the case without orientifold projection, we can pull back the $\T$ action to 
an action on $\Ms^\Sigma_d$. We will denote the $\T$ invariant subspace as 
$^\T\Ms^\Sigma_d$. Due to our restriction to homologically non-trivial boundaries, 
we have that $\Sigma$ can only be the union of $n$-pointed irreducible genus $g$ curves 
mapping under $f$ to one of the two non-invariant torus fixed points $p_1, p_2$, and
joined together by either 2-spheres or Klein handles. Furthermore, irreducible disk 
or cross-cap components can be attached to a contracted component. As before, it 
follows that each map $f$ can be represented combinatorially as a connected graph 
$\G$, with a bit of additional decoration. 

The contracted component curves correspond again to vertices decorated with the genus 
of each component, as well as by the point it maps to in target space. 
As before, the 2-spheres joining the contracted components are mapped to edges 
connecting the corresponding vertices. As a novelty, the Klein handles joining contracted
components are identified with Klein edges, which we may draw as an edge with a cross on 
top. Note that a Klein edge can be attached to a single vertex, \ie, it may form a 
loop (in distinction to an ordinary edge). We will refer to these Klein edges 
also as external Klein edges, while the Klein edges connecting two distinct vertices 
will be refered to as internal Klein edges. The disks and the cross-caps map to half-edges 
(also known as legs), or cross-edges attached to the vertices corresponding 
to the contracted component to which the disk or cross-cap are attached to, 
respectively. We will draw these simply as half-edges or half-edges with an arrow,
attached to vertices (with $i(v)=1$ or $2$ decoration). Note that there is a 
non-trivial restriction on graphs with Klein edges. Namely, since a 
Klein edge represents a handle (with orientation reversal), a proper graph 
should not split into disconnected components after removal of a Klein edge. 

As in the unorientifolded theory, each vertex can be associated to an ordinary 
moduli space $\Ms_{g_v,val(v)}$, such that the full graph corresponds to the moduli 
space
\beq
\Ms_{\G}=\prod_v \Ms_{g_v,val(v)}~.
\eq
Again, there is a morphism $\gamma_\G:\Ms_{\G}\rightarrow M_d^\G\subset\;
{}^\T\Ms_d^\Sigma$, which becomes an isomorphism if we quotient by $\A_\G$, 
the automorphism group of $\Ms_\G$. Thus,
\beq
\eqlabel{oloceq4}
^\T\Ms^{\Sigma}_d\cong\bigcup_{\G}\left(\Ms_{\G}/{ \A_{\G}}\right)~.
\eq
However, one has to be extra careful with $\A_\G$. In order to illustrate why, let
us slightly change our point of view.

To each curve $\Sigma$ we can associate a corresponding covering curve $\h\Sigma$ 
with $\Sigma=\h\Sigma/\sigma$. The covering space curve $\h\Sigma$ has genus 
$\h g=\chi+1$. Moreover each map $f$ can be lifted to a covering space map $\h f$ 
which is equivariant:
\beq
\eqlabel{oloceq2}
\h f=I\circ \h f \circ\sigma^{-1}~.
\eq
That is, the following diagram commutes:
\beq\label{oloceq1}
\begin{CD} 
\h \Sigma @>\h f>> X\\ 
@V{\sigma}VV @VV{I}V\\ 
\h\Sigma @> \h f>> X
\end{CD} 
\eq
Thus,  $\Ms^{\Sigma}_d$ can as well be defined as the fixed locus of the moduli space 
$\Ms^{\h \Sigma}_d$ of the corresponding doubled curve, \ie, $\Ms^{\Sigma}_d
={}^{\omega^*}\Ms^{\h\Sigma}_d$, with $\omega^*$ the map obtained by conjugating
with $I$ and $\sigma$, as in \eqref{oloceq2}. In particular,
\beq
^\T\Ms^{\Sigma}_d={}^{\omega^*\; {\h \T}}\Ms^{\h\Sigma}_d~.
\eq
Recall that to each $\h f\in {^{\h \T}}\Ms^{\h\Sigma}_d$ and $ f\in 
{^{ \T}}\Ms^{\Sigma}_d$ we have associated a corresponding graph $\h\G$, or $\G$,
respectively. In thinking about these various identifications, and their automorphism
groups, one's first naive expectation is that
\beq
\eqlabel{oloceq5}
\G=\h\G/\o^*~,
\eq
holds, with
\beq
\eqlabel{oloceq3}
|{\rm Aut}(\G)|= |{\rm Aut}(\h \G)^*|~,
\eq 
where ${\rm Aut}(\h \G)^*$ is the subgroup of ${\rm Aut}(\h \G)$ that commutes with $\o^*$. 
Note that $\o^*$ acting on $\Gamma$ leaves no vertices fixed, due to our restriction
to non-trivial boundaries and cross-caps.

To see that the relation is more subtle than described in \eqref{oloceq5} and 
\eqref{oloceq3}, note first that the inverse of relation (\ref{oloceq5}) is always true.
Namely, to a given graph $\G$ we can associate a corresponding covering space graph 
$\h\G$ via the following ``doubling" procedure: For each vertex $v$ draw a corresponding 
mirror vertex $v'$ with same $v(g)$ but mirror $i(v)$ decoration and for each edge 
draw a corresponding mirror edge. Then, for each disk and cross-cap connected to a 
vertex, draw an edge connecting the vertex with its mirror. Further, for each external 
Klein edge draw two edges connecting the vertex and its mirror, while for each internal 
Klein edge connecting the vertices $v_1$ and $v_2$ draw an edge connecting $v_1$ to 
$v_2'$ and one connecting $v_2$ to $v_1'$, where $v_i'$ are the mirror vertices.

\begin{figure}
\psfrag{I}[cc][][1]{$\cong$}
\psfrag{J}[cc][][1]{$\ncong$}
\psfrag{1}[cc][][1]{$\h\G_1$}
\psfrag{4}[cc][][1]{$\h\G_2$}
\psfrag{2}[cc][][1]{$\G_1$}
\psfrag{3}[cc][][1]{$\G_2$}
\psfrag{a}[cc][][0.75]{$\o^*$}
\psfrag{b}[ll][][0.75]{$\h\G_1/\o^*$}
\psfrag{c}[rr][][0.75]{$\h\G_2/\o^*$}
\begin{center}
\includegraphics[scale=0.3]{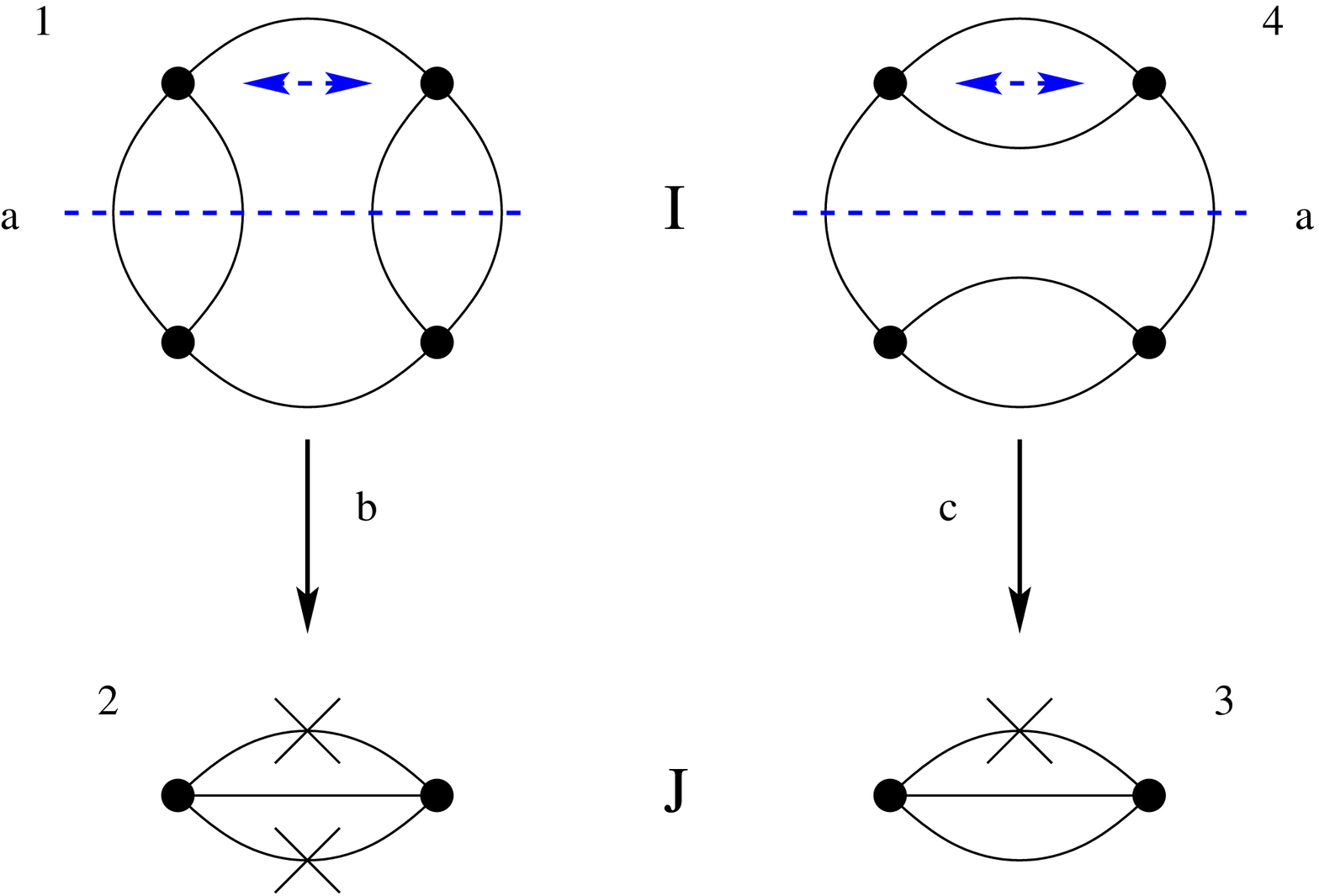}
\caption{The two graphs $\G_1\ncong\G_2$ can potentially contribute to $\t n_6^{(2,0)_k}$.
 However, we have that $\h\G_1\cong \h\G_2$, with $\G_1=\h\G_1/\o^*$ and 
$\G_2=\h\G_2/\o^*$, hence only one should contribute to  $\t n_6^{(2,0)_k}$.}
\label{olocfig2}
\end{center}
\end{figure}

However, while this doubling procedure gives a well-defined map $\G\mapsto\h\G$, 
there is generally no good inverse, \ie, relation (\ref{oloceq5}) does not 
hold in general. For example, consider the graphs $\h\G_1$ and $\h \G_2$ shown in 
figure \ref{olocfig2}. 
Both belong to the same equivalence class $[\h\G]$, \ie, there exists an isomorphism 
$a:\h\G_1\rightarrow\h\G_2$, equivariant with respect to $\omega^*$. However, the 
corresponding quotient graphs under $\o^*$ are not isomorphic. This is because in 
general the quotient graph $[\h \G]/\o^*$ depends on the choice of representative 
of $[\h\G]$, \ie, we have that
\beq
\eqlabel{oloceq3a}
[\h\G]/\o^*=\bigcup_i~ [\G_i],
\eq
where $[\G_i$] are equivalence classes of non-isomorphic quotient graphs $\G_i$. 
Nevertheless, the equivariance condition for $\h f$ implies that we should include 
only one graph $\G\in\{\G_i\}$, since $\h f$ should descend to a unique $f$. 

Hence, the relations $M_d^\G={}^{\o^*}M_d^{\h\G}\subset\;{}^\T\Ms_d^{\Sigma}$, and 
\eqref{oloceq5}, should be understood in the sense that they may include a choice 
of representative of $[\h\G]$. However, note that independent of a choice of 
representative, we have
\beq
\Ms_\G={}^{\omega^*}\Ms_{\h\G}=\sqrt{\Ms_{\h\G}}~.
\eq
The lesson we learn is the following. In order to avoid multiple countings we have 
to include in \eqref{oloceq4} only one representative of $\h\G/\o^*$. In practice, 
this means that we have to perform an extended isomorphism test on the set of graphs 
$\{\G\}$, \ie, two graphs need to be considered as identical if they are firstly
isomorphic after replacement of Klein edges with normal edges or if they secondly
lift to the same covering graph. 

\begin{figure}
\psfrag{a}[cc][][0.5]{$\o_1^*$}
\psfrag{b}[cc][][0.5]{$a$}
\psfrag{A}[ll][][0.75]{$\h\G/\o^*_1$}
\psfrag{B}[ll][][0.75]{$\h\G/\o^*_2$}
\psfrag{1}[cc][][1]{$\h\G$}
\psfrag{2}[cc][][1]{$\G_1$}
\psfrag{3}[cc][][1]{$\G_2$}
\psfrag{C}[ll][][0.75]{$\mathbf{12}$}
\psfrag{D}[ll][][0.75]{$\mathbf{6^*}$}
\psfrag{E}[ll][][0.75]{$\mathbf{2^*}$}
\begin{center}
\includegraphics[scale=0.3]{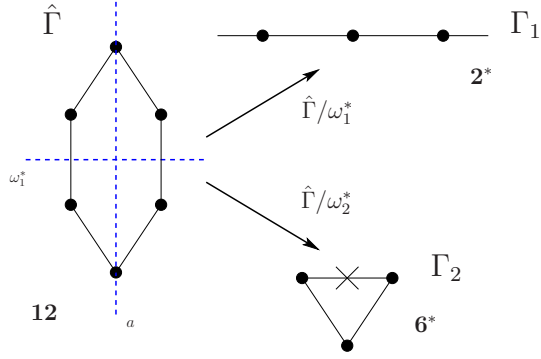}
\caption{The cyclic graph $\h\G=C_6$ with two differently acting involutions 
$\omega^*_i$. The involution $\o^*_1$ yields a quotient graph $\G_1$ with two 
half-edges contributing to $\t n_6^{(0,2)}$, while the involution $\o^*_2=(a\o^*_1)$ 
results in a graph $\G_2$ with a Klein edge contributing to $\t n_6^{(1,0)_k}$. The 
bold-face number is $|{\rm Aut}(\h \G)|$, while the bold-face numbers with star are 
the orders of the subgroups of ${\rm Aut}(\h \G)$ that commute with $\o^*_i$.}
\label{olocfig3}
\end{center}
\end{figure}

Let us now take a closer look at the relation \eqref{oloceq3}. As an illustrative 
example, consider the graph $\h \G$ with the two differently acting projections 
$\o^*_i$ sketched in figure \ref{olocfig3}.  
We see that $\o_1^*$ satisfies condition \eqref{oloceq3}, while $\o_2^*$ not. This 
raises the question whether $\A_\G$ involves ${\rm Aut}(\G)$ or ${\rm Aut}(\h \G)^*$. 
Again, the equivariance condition implies that ${\rm Aut}(\h \G)^*$ is the correct choice.
Hence,
\beq
\A_{\G}={\rm Aut}(\h {\G})^*\ltimes\left(\prod_c{\Z}_{d_c}\prod_e\Z_{d_e}\prod_k\Z_{d_k}\prod_h
\Z_{d_h}\right)~,
\eq 
where $k$ runs over the set of Klein edges, $h$ the set of half-edges and $c$ the set 
of cross-caps, if present.

Finally, incorporating the tadpole condition of \cite{Walcher:2007qp}, which tells 
us that graphs involving disks with even degree cancel against graphs with cross-caps, 
we deduce that the set $\{\G\}$ contributing to $\t n_d^{(g,h)}$ and $\t n_d^{(g,h)_k}$ 
includes all non-isomorphic and extended-non-isomorphic graphs $\G$ which fulfill 
the following criteria:
\begin{itemize}
\setlength{\itemsep}{0.5pt} 
\item $d_h$ is odd for all half-edges.
\item $2\sum_e d_e+2\sum_k d_k+\sum_h d_h=d$.
\item $1-2|v|+2|e|+2|k|+|h|+2\sum_v g_v=g$, where $|k|$ is the number of Klein edges 
and $|h|$ the number of half-edges.
\item Edges connect only vertices with $i(e)\neq j(e)$.
\item Half-edges are only attached to vertices with $i(v)=1$ or $2$.
\item Klein edges only connect vertices with $i(k)=j(k)$ or with $i(k)=1$ or $2$ and 
$j(k)=3$ or vice-versa.
\end{itemize}

Then, with $\G=\h\G/\o^*$ we obtain from (\ref{loceq2}):
\beq
\eqlabel{oloceq6}
\t n^{\Sigma}_d=(-1)^{3g-3+h}\sum_{\G}\frac{(-1)^{|k|}}{|\A_{\G}|}\int_{\Ms_{\G}}
\sqrt{\frac{\mathbf{e}(i^*\E_d)}{\mathbf{e}(\N_{\h\G}^{vir})}}~,
\eq
where the sum runs over the set $\{\G\}$ specified above. Note that our discussion 
does not a priori fix the overall sign nor the sign of each individual graph. 
However, most of the sign factors in \eqref{oloceq6} can actually be borrowed from
the tree-level discussion in \cite{PSW}. The remaining signs were determined in
\cite{Walcher:2007qp} based on computations on compact models, comparison with
the B-model, and integrality of Gopakumar-Vafa invariants. The existence of the 
sign $(-1)^k$ can also be inferred from the requirement that the contribution
of a given class of equivariant graphs should be independent of the chosen
quotient representative, see discussion around eq.\ \eqref{oloceq3a}.

The contribution of vertices, edges and Klein edges of the quotient space graph $\G$ 
to the integrand of (\ref{oloceq6}) is as before accounted for by (\ref{loceq3}) and 
\eqref{loceq4}, and supplemented by the following modifications. For each half-edge 
ending on a vertex $v$, add a flag $(v,h)$ to the set of flags of $v$. 
Define $i(h)$ as the image point $p_i$ to which $v$ maps in target space and $j(h)$ 
the image point $p_j$ of the corresponding mirror-vertex in the covering graph.
We also multiply the integrand by the following factor accounting for the half-edges.
(This is essentially just a squareroot of an ordinary edge contribution.)
\beq
\begin{split}
D(\G)=&\prod_h \frac{(-1)^{\frac{d_h-1}{2}}d_h^{d_h}}{(d_h!)(\lambda_{i(h)}-
\lambda_{j(h)})^{d_h}}\prod^{\frac{d_h-1}{2}}_{\substack{k\neq i(h),j(h)\\a=0}}
\frac{1}{\frac{a}{d_h}\lambda_{i(h)}+ \frac{d_h-a}{d_h}\lambda_{j(h)}-\lambda_k}\\
&\times \prod_h^{\frac{3 d_h-1}{2}}\left[\Lambda_{i(h)}+\frac{m}{d_h}(\lambda_{i(h)}
-\lambda_{j(h)})\right]~.
\end{split}
\eq
The Klein edges are treated like usual edges, however with $i(k)$ and $j(k)$ defined 
as $i(v)$ and $j(v)$ of the corresponding covering graph edge. At the very end, we
need to identify in the integrand $\lambda_1=-\lambda_2$. Then we cancel any 
common factors between numerator and denominator from each summand. These could
cause ill-defined ``$\frac 00$''-type expressions when we set $\lambda_3=0$
in the final expression for $\t n_d^\Sigma$.

We have developed a full computer implementation of the above prescription 
and used it to calculate the open and unoriented Gromov-Witten invariants up to 
$\chi=9$ for various degrees. We will not list the complete data, but rather just 
give the real Gopakumar-Vafa invariants which we were able to verify with our data,
see appendix \ref{secA}. Some of the Gromov-Witten invariants that we obtained can 
be inferred from the large-volume expansions given in section \ref{fixholamb}. These 
amplitudes were computed by using our localization data to fix the holomorphic
ambiguities of the B-model. This will be explained in detail in section \ref{Bmodel}.

\section{The real topological vertex}
\label{realvertex}

The localization computations of the previous section quickly become rather complicated 
with increasing genus and degree. There are two sources of complexity. First, one has to
generate the decorated graphs and correctly determine their automorphism groups. As we have
seen, this can be tricky especially in the real case. Second, one has to evaluate the
graphs, and in particular to compute the Hodge integrals. The best available general
algorithm for this still is Faber's. On the other hand, note that the computation of the
Hodge integral is a local problem, attached to the fixed points of the torus action.
Some years ago, it has been realized that there is in fact a closed formula that resums 
the requisite Hodge integrals to all orders in the genus expansion, and that incidentally
also solves the first-mentioned graph combinatorial problem in a very efficient way.
This is the topological vertex \cite{Aganagic:2003db}.

\subsection{Topological vertex for local \texorpdfstring{$\P^2$}{P2}}

Instead of setting up the full formalism of \cite{Aganagic:2003db}, we give here an 
elementary account of the topological vertex at work on local $\P^2$. This will be 
sufficient to write down the formulas that compute the amplitudes also in the real case. 

\begin{figure}
\psfrag{R1}{$R_1$}
\psfrag{R2}{$R_2$}
\psfrag{R3}{$R_3$}
\begin{center}
\epsfig{height=4cm,file=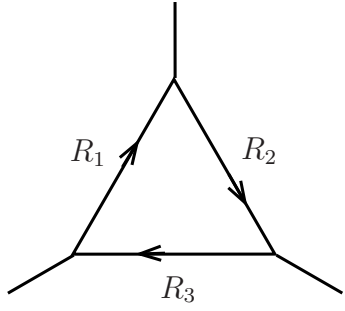}
\end{center}
\caption{Trivalent diagram representing local $\P^2$ for the purposes of evaluating
the topological vertex.}
\label{toricdi}
\end{figure}

The toric diagram representing local $\P^2$ as a $T^2\times\reals$ fibration over a
three-dimensional base is shown in figure \ref{toricdi}.
According to \cite{allloop,iqbal}, the total closed topological string partition 
function of local $\P^2$ is given by
\begin{equation}
\eqlabel{givenby}
Z = \sum_{R_1,R_2,R_3} (-1)^{\sum l(R_i)} q^{-\sum\kappa_{R_i}}
\ee^{-t \sum l(R_i)}
C_{0R_3^t R_1}C_{0R_2^tR_3} C_{0 R_1^t R_2}~.
\end{equation}
In this sum, the $R_i$ run over all Young diagrams (representations of $U(\infty)$),
$l(R_i)$ is the number of boxes in $R_i$, and $\kappa(R_i)$ is related to the second
Casimir of the corresponding representation. These initial factors come from the need 
to adjust the framing on the internal legs between the vertices. But the central 
ingredient of \eqref{givenby} is the topological 
vertex itself. The full three-legged vertex (in the canonical framing) is given by
\begin{equation}
\eqlabel{vertex}
C_{R_1 R_2 R_3} = 
q^{\kappa_{R_2}/2+\kappa_{R_3}/2}
\sum_{Q,Q_1,Q_3} N_{Q Q_1}^{R_1} N_{Q Q_3^t}^{R_3^t} 
\frac{W_{R_2^t Q_1} W_{R_2 Q_3^t}}
{W_{R_20}}~,
\end{equation}
where the $N_{Q Q_1}^{R_1}$ are the $U(\infty)$ tensor product coefficients, and 
$W_{R_1R_2}=W_{R_1R_2}(q)$ is a certain rational function of $q$ that arises by 
taking a specific limit (in level and rank) of the Chern-Simons invariant of the 
Hopf link in $S^3$ decorated with $R_1$ and $R_2$. When one of the representations 
on the vertex is trivial, we have the more compact expression
\begin{equation}
C_{0R_1R_2} = q^{\kappa_{R_2}/2} W_{R_1 R_2}~.
\end{equation}
In all these formulas, we have adopted the standard topological vertex notation in which
$t$ still denotes the K\"ahler parameter of $\P^2$, but $q=\ee^{\lambda}$ is the 
exponentiated string coupling. We make full contact with the previous notation by 
relating the free energy in those variables to the Gopakumar-Vafa invariants (cf. 
(\ref{standardgv}))
\begin{equation}
\eqlabel{contact}
\F = \log Z = \sum_{d,g,k} N_{d}^{(g)} 
\frac 1k\bigl(q^{k/2}-q^{-k/2}\bigr)^{2g-2} \ee^{-tkd}~.
\end{equation}

\subsection{Taking a squareroot}

We are now in a position to present the formulas that express the real topological
string amplitudes of local $\P^2$ in terms of the (real) topological vertex. 
The basic idea is the following. The topological vertex can be viewed as an 
all-genus resummation of the local contribution at each vertex on the toric 
diagram to the localization formulas for the topological string amplitude (see,
\eg, \cite{diaconescu}). Going from the ordinary topological string to the real 
topological string amounts in the localization formalism to first restrict to the 
graphs fixed under the target space involution, and then take a squareroot of each 
individual contribution. The only conceptual difficulty is to understand which sign 
of the squareroot to take. 

Taking these observations together, all we have to do to obtain a real vertex formalism
is to identify the action of the target space involution on the toric diagram of figure 
\ref{toricdi} and on formulas \eqref{givenby} and \eqref{vertex}, and then to take an 
appropriate squareroot. It is in fact not hard to see that the action on the 
representations is $R_1\mapsto R_2$ and $R_3\mapsto R_3$. Using the symmetry of the 
topological vertex 
\begin{equation}
C_{R_1R_2R_3} = q^{\sum \kappa_{R_i}/2} C_{R_1^tR_3^t R_2^t}~,
\end{equation}
we see that for the fixed configurations, $R_1=R_2$, the summand in \eqref{givenby} is 
of the form.
\begin{equation}
(-1)^{2l(R_1)+l(R_3)} q^{-5\kappa_{R_1}/2-\kappa_{R_3}/2}\ee^{-t(2l(R_1)+l(R_3))}
\bigl(C_{0R_3^tR_1}\bigr)^2 C_{0 R_1^t R_1}~.
\end{equation}
This is a perfect square except for the final term, which arises at the vertex fixed
under the involution. Such a term will arise in general toric Calabi-Yaus with involution
that leaves some vertices fixed, but permutes two of the legs ending on it. In that
case, we will generally require a ``real topological vertex'' that might be obtained
by taking an appropriate squareroot of the expression \eqref{vertex} for the topological
vertex with $R_3=R_1^t$, and $R_2=R_2^t$. Indeed, we see that with this external data,
and restriction of the sum to $Q_3=Q_1^t$, the vertex is itself almost a sum of squares,
\begin{equation}
\bigl(N_{Q Q_1}^{R_1}\bigr)^2 \frac{\bigl(W_{R_2^t Q_1}\bigr)^2}{W_{R_20}}\,,
\end{equation}
except for the $W_{R_20}$ in the denominator. We do not know at present how to
take a squareroot of that last term. But luckily, for our application to local
$\P^2$, we only need the two-legged vertex, and the real vertex only with trivial
representation $R_2=0$ on the fixed leg. Based on the above observations, we propose
the following expression for that real vertex amplitude
\begin{equation}
\eqlabel{realvert}
C^{\rm real}_{R_1 0} = q^{-\kappa_{R_1}/4} \sum_{Q,Q_1} N_{QQ_1}^{R_1} W_{Q_10}~.
\end{equation}
Returning to the formula for local $\P^2$, we obtain for the partition function
of the real topological string
\begin{equation}
Z^{\rm real} =\sum_{R_1,R_3} (-1)^{l(R_1)} (-1)^{p(R_3)} 
\ee^{-t(l(R_1)+l(R_3)/2)}
q^{-5\kappa_{R_1}/4 -\kappa_{R_3}/4} C^{\rm real}_{R_10} C_{0R_3^tR_1}~,
\end{equation}
where $(-1)^{p(R_3)}=\pm 1$ is an a priori undetermined sign. Note that for 
symmetry reasons, this sign can only depend on $R_3$, as we have indicated. Some 
experimentation shows that its correct value is determined by the {\it number 
of boxes in even columns}. In other words, if $R_3^t$ consists of rows of length 
$l_1,\ldots l_r$, then
\begin{equation}
\eqlabel{weird}
(-1)^{p(R_3)} = (-1)^{\sum_{i} l_{2i}}~.
\end{equation}
We are not aware that such a sign associated with 2d partitions has appeared before, 
nor does there seem to be any representation theoretic meaning. This would be worthy
of clarification. 

In any event, we can now make contact with the other expressions for the amplitudes of
real local $\P^2$. The real analogue of \eqref{contact}, see also \eqref{possesses}, is
\begin{equation}
\log Z^{\rm real} = \frac 12 \F + 
\sum_{\topa{d\equiv\chi \bmod 2}{k\;{\rm odd}}} N'^{(\chi)}_d \frac 1k 
\bigl(q^{k/2}-q^{-k/2}\bigr)^\chi\ee^{-t k d/2}\epsilon^\chi~.
\end{equation}
These formulas reproduce the localization results of the previous section, wherever
the available data has allowed comparison, and also agree with the developments of the 
B-model to which we turn presently.

To close this section, we point out that we have merely scratched the surface of the
real topological vertex. Starting with the derivation, but including its properties, 
applications, and connections with other theories, one can ask for a real counterpart
of essentially everything that is known about the ordinary topological vertex. The central
question in this endeavour is whether the signs can be understood in a uniform way.
We have to leave this for the future.

\section{The B-model}
\label{Bmodel}

We now turn to a computation of the real topological string amplitudes using the 
mirror B-model. Here again, most of the technology is already in place in the literature,
so we will be rather brief, and just restate the formulas in our chosen normalization.
The main aim is to push the holomorphic anomaly technique to higher order in perturbation
theory. Besides reproducing the A-model results from section \ref{loc} and the results from 
the real topological vertex from section \ref{realvertex}, the main payoff will be a new 
gap structure in the expansion of the real topological string amplitudes at the conifold.

In order to set the stage, let us briefly recall some basic facts. While we introduced 
the A-model topological string free energies $\F(t),\K(t)$ and $\mathcal G(t)$ 
in a rather geometric way as a count of holomorphic maps from world-sheets 
with specific topology into a Calabi-Yau manifold with K\"ahler parameter $t$, it is 
important to note that this interpretation only holds at large volume. Away from this 
point in moduli space, classical notions of geometry break down and so does the original 
interpretation of the free energies. On the other hand, the proper definition of 
the perturbative amplitudes is really in terms of the topologically twisted 2d world-sheet 
theory, which is well-defined over the entire stringy K\"ahler moduli space. Here it 
is where mirror symmetry comes to rescue, since the A-twisted world-sheet theory on 
$X$ is equivalent to a B-twisted theory on a mirror Calabi-Yau geometry $Y$, with K\"ahler 
parameter traded for complex structure, such that the B-model captures the quantum 
regime of the A-model. In particular, the corresponding B-model amplitudes
$\F(z,\b z),\K(z,\b z)$ and $\mathcal G(z,\b z)$ are now functions 
over the complex structure moduli space of $Y$, which we will denote 
as $\mathcal M_Y$. 

A key point that allows to efficiently solve for the amplitudes in the B-model 
is that their anti-holomorphic derivatives over $\mathcal M_Y$ do not vanish 
\cite{Bershadsky:1993cx}, as we have indicated in
the notation. The so-called holomorphic anomaly equations \cite{Bershadsky:1993cx,
Walcher:2007tp,Walcher:2007qp}, completely determine the anti-holomorphic dependence 
of the amplitudes, and reduce the problem to the fixing of the holomorphic part. 
Constraints of modular invariance and a priori knowledge about the compactification of the
moduli space make this a finite-dimensional problem. Its general solution is
still rather elusive, but important progress has been made in recent years.
An additional bonus of the B-model is the possibility to analyze the structure of
the amplitudes at special points in moduli space other than large volume.

\subsection{Solving the (extended) holomorphic anomaly equations}

The extended holomorphic anomaly equations of \cite{Walcher:2007tp,Walcher:2007qp}, 
specialized to the local 1-parameter case, are given by 
\beq
\eqlabel{han1}
\d_{\b z}\F^{(g,h)}=\frac{1}{2}\sum_{\substack{g_1+g_2=g\\h_1+h_2=h\\2g_i+h_i>1}}
C^{zz}_{\b z}\F_z^{(g_1,h_1)}\F_z^{(g_2,h_2)}+\frac{1}{2}C^{zz}_{\b z}\F_{zz}^{(g-1,h)}-
\Delta^z_{\b z}\F_z^{(g,h-1)}~,
\eq
and
\beq
\eqlabel{han2}
\begin{split}
\d_{\b z}\K^{(g,h)}=&\sum_{\substack{g_1+g_2=g\\h_1+h_2=h\\2g_2+h_2>1\\g_1>0}} 
C^{zz}_{\b z}\K_z^{(g_1,h_1)}\F_z^{(g_2,h_2)}+\frac{1}{2} 
\sum_{\substack{g_1+g_2=g\\h_1+h_2=h\\g_i>0}} C^{zz}_{\b z}\K_z^{(g_1,h_1)}
\K_z^{(g_2,h_2)} \\
&+ C^{zz}_{\b z}\K^{(g-1,h)}_{zz} +\frac{1}{2}C^{zz}_{\b z}\F_{zz}^{(g-1,h)}-
\Delta^z_{\b z}\K_z^{(g,h-1)},
\end{split}
\eq
where $\F_{z\cdots z} = D_z\cdots D_z\F$, similarly for the $\K$, and $z$ is a local
coordinate on the space of complex structures, $\mathcal M_Y$, of $Y$. Further, 
$C_{zzz}$ is the usual Yukawa-coupling, \ie, the sphere three-point function, and 
$\D_{zz}$ is the disk two-point function (with bulk insertions). As usual, indices 
are raised and lowered via the K\"ahler metric on $\mathcal M_Y$. Note that we have 
here already implemented tadpole cancellation, so we can consistently set the
$\R^{(g,h)}$ to zero.

Equations \eqref{han1} and \eqref{han2} can be solved recursively. Let us for the moment 
consider the simplified case without open strings, \ie, $h=0$. Then, recursively solved, 
the equations give an expression for $\F^{(g,0)}$ and $\K^{(g,0)}$ in terms of 
$\F^{(1,0)}$ and $\K^{(1,0)}$. These $1$-loop amplitudes have the following holomorphic
limits \cite{Bershadsky:1993cx,Walcher:2007qp}
\beq
\eqlabel{oneloop}
\begin{split}
\F^{(1,0)}&=\frac{1}{2}\log{(\tau)}+a_\F^{(1,0)}~,\\
\K^{(1,0)}&=\frac{1}{2}\log{(\tau)}+a_\K^{(1,0)}~,
\end{split}
\eq
where we defined $\tau=\d_t z(t)$ to be the derivative of $z$ with respect to the 
preferred flat coordinate $t$ at the large volume point of $\mathcal M_Y$. 
The 1-loop holomorphic ambiguities occurring in (\ref{oneloop}) are for local 
$\P^2$ given by 
\beq
\begin{split}
a_\F^{(1,0)}&=-\frac{1}{2}\log(z)-\frac{1}{12}\log(-z)-\frac{1}{12}\log(1-27z)~,\\
a_\K^{(1,0)}&=-\frac{1}{2}\log(z)- \frac{1}{8}\log(1-27z)~.
\end{split}
\eq
To proceed, we define the non-holomorphic objects 
(propagators 
in Feynman diagram language) $S^{zz}$ and $K^{zz}$ as
\beq
\eqlabel{Bholeq1}
\begin{split}
S^{zz}&=2~\frac{\F_z^{(1,0)}}{C_{zzz}}~,\\
K^{zz}&=2~\frac{\K_z^{(1,0)}}{C_{zzz}}~,\\
\end{split}
\eq
where the Yukawa coupling $C_{zzz}=\F^{(0,0)}_{zzz}$ reads for local $\P^2$
\beq
C_{zzz}=-\frac{1}{3}\frac{1}{z^3(1-27z)}~.
\eq
Comparing with \eqref{oneloop}, we see that $K^{zz}$ and $S^{zz}$ differ only by a 
holomorphic function
\beq
 K^{zz}=S^{zz}+2\frac{\d_z a_{\K\F}}{C_{zzz}}~,
\eq
with
\beq
 a_{\K\F}=a_\K^{(1,0)}-a_\F^{(1,0)}~.
\eq
Hence, we can express both $\F_z^{(1,0)}$ and $\K_z^{(1,0)}$ in terms of the single 
non-holomorphic propagator $S^{zz}$, up to holomorphic terms. Furthermore, (using 
the special geometry relation) it is easy to deduce that one can re-express the 
covariant derivative of $S^{zz}$ in terms of $S^{zz}$, \ie,
\beq
D_z S^{zz}=-C_{zzz} \left(S^{zz}\right)^2+(a_{DS})^{zz}_z~.
\eq
and that a similar condition holds for the 
connection coefficient $\G^z_{zz}$,
\beq
\G^z_{zz}=-C_{zzz}S^{zz}+(a_\G)^z_{zz}~.
\eq
Here, $a_{DS}$ and $a_{\Gamma}$ are global holomorphic functions. For local
$\P^2$, and our definition of the propagator \eqref{Bholeq1}, we have
\begin{gather}
a_\G=2~\d_za_\F^{(1,0)}=-\frac{7-216z}{6z(1-27z)}~, \\
a_{DS}=-\frac{z}{12(1-27z)}~.
\end{gather}
Thus, we conclude that all $\F^{(g,0)}$ and $\K^{(g,0)}$ can be expressed as 
polynomials in the single propagator $S^{zz}$, with coefficients given by 
holomorphic functions in $z$. This idea originated in \cite{Yamaguchi:2004bt}, 
to which we refer for more details about the $\F^{(g,0)}$ case.

Let us now include the open string sector. With assumptions detailed in 
\cite{Walcher:2007tp}, the only new ingredient that enters the recursive solution
is the disk amplitude with two bulk insertions. In the holomorphic limit, this is 
given by \cite{Walcher:2007tp}
\beq
\D_{zz}=\F_{zz}^{(0,1)}=\d_z\d_z\mathcal T,
\eq
where $\mathcal T$ is the domain-wall tension. As for the closed string case, we can 
define a non-holomorphic object (terminator in Feynman diagram language)
\beq\label{Bholeq3}
\Delta^z=-\frac{\F_{zz}^{(0,1)}}{C_{zzz}}~,
\eq
which for local $\P^2$ satisfies
\beq
D_z\Delta^z=\frac{3}{4}\sqrt{z}~.
\eq
As a consequence, the amplitudes $\F^{(g,h)}$ and $\K^{(g,h)}$ can be expressed in terms 
of the two non-holomorphic objects $S^{zz}$ and $\D^{z}$, with holomorphic coefficients. 
A detailed discussion of the (oriented) $\F^{(g,h)}$ case can be found in 
\cite{Konishi:2007,Alim:2007qj}. Then, using the relations \cite{Bershadsky:1993cx,
Walcher:2007tp}
\beq
C^{zz}_{\b z}=\d_{\b z}S^{zz}, ~\Delta^z_{\b z}=\d_{\b z} \Delta^z,
\eq
one can re-express the above extended holomorphic anomaly equations as 
\beq
\begin{split}
\d_{S^{zz}}\F^{(g,h)}&=\frac{1}{2}\sum\F_z^{(g_1,h_1)}\F_z^{(g_2,h_2)}+
\frac{1}{2}\F_{zz}^{(g-1,h)}~,\\
\d_{\Delta^z}\F^{(g,h)}&=-\F_z^{(g,h-1)}~,
\end{split}
\eq
and
\beq
\begin{split}
\d_{S^{zz}}\K^{(g,h)}&=\sum\K_z^{(g_1,h_1)}\F_z^{(g_2,h_2)}+\frac{1}{2}
\sum\K_z^{(g_1,h_1)}\K_z^{(g_2,h_2)}+\K_{zz}^{(g-1,h)}+\frac{1}{2}\F_{zz}^{(g-1,h)}~,\\
\d_{\Delta^z}\K^{(g,h)}&=-\K_z^{(g,h-1)}~,
\end{split}
\eq
These equations can be easily solved by direct integration, up to the holomorphic 
ambiguities to which we will return momentarily. 

Before that, recall that in \eqref{inteq0a} we have identified the total topological string 
amplitude $\Gx$ as a combination of $\F$'s and $\K$'s (see \eqref{inteq2}, with 
$\R^{(g,h)}\equiv 0$). It is clear that one can write down a combined holomorphic 
anomaly eqation directly for the total amplitude $\Gx$, which is in fact somewhat 
simpler \cite{Walcher:2007qp} (as already stressed in the introduction, we are 
working here with a different normalization of $\Gx$, as a result the combined 
anomaly equation we are using differs slightly from the one presented in 
\cite{Walcher:2007qp})
\beq
\eqlabel{Bholeq2}
\d_{\b z}\Gx=\frac12\sum_{\substack{\chi_1+\chi_2=\chi-2\\ \chi_i\geq 0}}C^{zz}_{\b z}
\Gx_z \Gx_z+C^{zz}_{\b z}\mathcal G^{(\chi-2)}_{zz}-\Delta^z_{\b z}\mathcal G_z^{(\chi-1)}~.
\eq
It is obvious that just as the individual amplitudes $\F$ and $\K$, $\Gx$ can be written 
as a polynomial in the non-holomorphic propagator $S^{zz}$ and terminator $\D^z$,
with holomorphic coefficients. Thus, we can re-express (\ref{Bholeq2}) as
\beq
\eqlabel{reexpress}
\begin{split}
\d_{S^{zz}}\Gx&=\frac12\sum\Gx_z \Gx_z+\mathcal G^{(\chi-2)}_{zz}~,\\
\d_{\Delta^z}\Gx&=-\mathcal G_z^{(\chi-1)}~,
\end{split}
\eq
which again can be simply solved by integration, yielding a polynomial in $S^{zz}$ and 
$\D^z$ with holomorphic functions in $z$ as coefficients.

\subsection{Fixing the holomorphic ambiguities}
\label{fixholamb}

In order to evaluate the polynomials in $S^{zz}$ and $\D^z$ that we have obtained by
integrating the holomorphic anomaly equation, \ie, to obtain explicit 
expansions of $\F$, $\K$ and $\mathcal G$, we have to specify the coordinate $z$.
That is, we have to chose a point in moduli-space around which to expand these amplitudes. 
Furthermore, the holomorphic ambiguities of these amplitudes, which we will denote as 
$a^{(g,h)}_{\F/\K}$ and $a^{(\chi)}_{\mathcal G}$, have to be fixed.

The natural point of interest in moduli space is the large-volume point with flat 
coordinate $t$ corresponding to the K\"ahler parameter of $\P^2$. At this point, 
we can compare with our results from localization and the real 
topological vertex to fix the ambiguities $a^{(g,h)}_{\F/\K}$ and $a^{(\chi)}_{\mathcal G}$. 
The mirror map $z(t)$ and the domain-wall tension $\mathcal T$ that enters into $\D^z(t)$ 
can be obtained from the (inhomogenous) Picard-Fuchs equation 
(we have taken the liberty to multiply the inhomogeneous part with an additional
factor of $-\ii(2\pi)^2$ in comparison with \cite{Walcher:2007qp})
\beq\label{faeq1}
(\th^3-3 z\th(3\th+1)(3\th+2))~{\mathcal T}=-\frac{1}{4}\sqrt{z}~,
\eq
with $\th=z\d_z$. The solutions of the homogenous equation near $z=0$ yield 
the well-known closed string periods (leading to the mirror map $z(t)$), while the 
solution of the inhomogeneous equation gives the domain-wall tension interpolating 
between the two open string vacua (recall that we have a discrete $\Z_2$ valued 
Wilson-line on the brane). 
\beq
\mathcal T=2\i ~\Gamma(3/2)^2\sum_{n=0}^\infty\frac{\Gamma(3n+3/2)}
{\Gamma(n+3/2)^3}z^{n+1/2}~.
\eq
Using the definitions \eqref{Bholeq1} and \eqref{Bholeq3}, we obtain the following 
large-volume expansions of the $z(t)$, $S^{zz}(t)$ and $\D^z(t)$
\begin{equation}
\begin{split}
z(t)&\textstyle =-q - 6 q^2 - 9 q^3 - 56 q^4 + 300 q^5 - 3942 q^6 + 48412 q^7 - 
\cdots ~,\\
\textstyle
 S^{zz}(t)&\textstyle =\frac{1}{2}q^2 + 15 q^3 + 135 q^4 + 785 q^5 + 
\frac{4473}{2} q^6 + 18333 q^7 - 
 \cdots ~,\\
\textstyle
-\i\D^{z}(t)&\textstyle =-\frac{3}{2} q^{3/2} - \frac{39}{2} q^{5/2} - 
 \frac{117}{2} q^{7/2} - \frac{765}{2} q^{9/2} + 
 1881 q^{11/2} -\cdots ~,
\end{split}
\end{equation}
with $q=e^{2\pi\ii t}$. Note that
\beq
S^{zz}(t)=\tau^2S^{tt},~S^{z}(t)=\tau\D^t~.
\eq
where $\tau=\del_t z(t)$. Plugging these expansions into the polynomial expressions 
for $\F$ and $\K$ and comparing with our localization results allows us to fix the
holomorphic ambiguities up to a certain order. We here report our observations.

First of all, the holomorphic ambiguities of $\F^{(0,h)}$, $\F^{(1,h)}$ and $\K^{(1,h)}$ 
take a very simple form. More precisely, in our scheme, the ambiguities 
$a_\F^{(0,h)}$ and $a_\K^{(1,h)}$ all vanish, whereas we find for the 
ambiguity $a_\F^{(1,h)}$ of $\F^{(1,h)}$
\beq
\eqlabel{canonical}
a_\F^{(1,h)}=
\left\{\begin{matrix}
-\frac{1}{24} z^{1/2}& h=1\\
 (-1)^h \frac{3^{(h-1)}}{2^{(2h+2)}h}~z^{h/2}&h>1
\end{matrix}
\right.~.
\eq
Secondly, one may note that the open string degenerations alone completely generate 
all Feynman diagrams for $\F^{(0,h)}$, $\F^{(1,h)}$, and $\K^{(1,h)}$ 
for all $h$. This means that using a flat coordinate $t$, we have the 
following simple expressions for these amplitudes, which can be evaluated even 
for very large $h$ most economically:  
\beq
\eqlabel{persist}
\begin{split}
\F^{(0,h)}&=\int d\Delta^t  \d_t \F^{(0,h-1)}=\left[\int d\Delta^t \d_t\right]^{h-2}
\F^{(0,2)}(t)~,\\
\K^{(1,h)}&=\int d\Delta^t  \d_t \K^{(1,h-1)}=\left[\int d\Delta^t \d_t\right]^{h}
\K^{(1,0)}(t)~,\\
\F^{(1,h)}&=\int d\Delta^t  \d_t \F^{(1,h-1)}+a_\F^{(1,h)}\\
&=\left[\int d\Delta^t \d_t\right]^{h}\F^{(1,0)}(t)+\sum_{i=1}^h~\left[\int d\Delta^t 
\d_t\right]^{(h-i)}a_\F^{(1,i)}~.\\
\end{split}
\eq
For higher genus, things become more involved, and there does not appear to be a
simple structure as in \eqref{canonical}. For illustration, we give
here the following oriented open string amplitudes
\begin{equation}
\begin{split}
\textstyle
\F^{(2,1)}&\textstyle =-\frac{7 \sqrt{q}}{2880}+\frac{79 q^{3/2}}{2880}-\frac{59 q^{5/2}}{128}+
\frac{2597 q^{7/2}}{720}-\frac{205151 q^{9/2}}{240}+\frac{31659529
   q^{11/2}}{640}+\cdots ~,\\
\textstyle
\F^{(2,2)}&\textstyle =\frac{11 q}{3072}+\frac{41 q^2}{12288}+\frac{10663 q^3}{2560}-
\frac{389561 q^4}{30720}+\frac{13173223 q^5}{3072}-\frac{5413756009
   q^6}{20480}+\cdots ~,\\
\textstyle
\F^{(2,3)}&\textstyle =-\frac{87 q^{3/2}}{20480}-\frac{3259 q^{5/2}}{10240}-
\frac{476291 q^{7/2}}{20480}-\frac{465417 q^{9/2}}{20480}-\frac{348949197
   q^{11/2}}{20480}+\cdots ~,\\
\textstyle
\F^{(2,4)}&\textstyle =\frac{407 q^2}{81920}+\frac{57861 q^3}{32768}+\frac{2103243 q^4}{20480}+
\frac{15796159 q^5}{32768}+\frac{4897896903 q^6}{81920}+\cdots ~.
\end{split}
\end{equation}
and the following unoriented amplitudes. 
\begin{equation}
\begin{split}
\textstyle
\K^{(2,0)}&\textstyle =\frac{5 q}{128}+\frac{33 q^2}{16}-\frac{10953 q^3}{64}+\frac{223495 
q^4}{32}-\frac{13926207 q^5}{64}+\frac{379810917 q^6}{64}+\cdots ~,\\
\textstyle
\K^{(2,1)}&\textstyle =-\frac{9 q^{3/2}}{128}-\frac{12723 q^{5/2}}{1024}+\frac{270585 
q^{7/2}}{256}-\frac{13282137 q^{9/2}}{256}+\frac{1951535727 q^{11/2}}{1024}+\cdots ~,\\
\textstyle
\K^{(2,2)}&\textstyle =\frac{99 q^2}{2048}+\frac{48897 q^3}{1024}-\frac{4235175 q^4}{1024}+
\frac{120073203 q^5}{512}-\frac{20153395269 q^6}{2048}+\cdots ~,\\
\textstyle
\K^{(2,3)}&\textstyle =\frac{747 q^{5/2}}{4096}-\frac{4921425 q^{7/2}}{32768}+\frac{215009073 
q^{9/2}}{16384}-\frac{27419944149 q^{11/2}}{32768}+\cdots ~,\\
\textstyle
\K^{(2,4)}&\textstyle =-\frac{34749 q^3}{32768}+\frac{6909435 q^4}{16384}-\frac{1208349657 
q^5}{32768}+\frac{21269586123 q^6}{8192}+\cdots ~.
\end{split}
\end{equation}
In all these cases, we have parameterized the holomorphic ambiguities of $\F^{(g,h)}$ 
and $\K^{(g,h)}$ via the function 
\beq\label{FKambi}
 a^{(g,h)}_{\F/\K}=\sum_{i=0}^{n-1} a_i\frac{z^{i+h/2}}{(1-27z)^{2g-2}}~,
 \eq
where $a_i$ are rational numbers and
\beq
n=
\left\{
\begin{matrix}
2g-1&\text{for}~\F^{(g,0)} \\
3g-2&\text{else}
\end{matrix}
\right.~.
\eq
We have then compared the coefficients of the $q$-expansion in low degree with 
our localization results in order to determine the coefficients of the holomorphic
ambiguity $a_i$. Note that the number of coefficients that needs to be fixed is larger
for $h\neq 0$ than in the purely closed string case. This can be traced back to the
existence of the tensionless domain wall at the orbifold point and the resulting 
singularity of the $\F$ and $\K$ at this point. On the other hand, it is mildly 
comforting that the number of unknown coefficients does not grow with $h$. (Naively,
one might expect $n\sim 3g+h$ or something similar.) This could suggest that there 
is additional structure that we have so far not identified. However, hopes of
finding a very simple expression as in \eqref{canonical} for $g>1$ have so far 
not materialized.

The (individual) amplitudes we have determined so far are only sufficient to obtain 
$\Gx$ via relation (\ref{inteq2}) up to $\chi=3$ (which has been already achieved 
in \cite{Walcher:2007qp}). In order to go beyond we need more information. A prime 
candidate to look at is the conifold point in moduli space, where 
it is known that the expansion of the closed string amplitudes $\F^{(g,0)}$ possesses 
a ``gap''. This structure, whose existence can be understood physically, gives 
enough information to completely determine these amplitudes for all $g$ 
\cite{Huang:2006si,Huang:2006hq,Haghighat:2008gw}. It is natural to ask whether 
there is as well some systematics in the expansion of the real topological string 
amplitudes at the conifold point.

To exhibit the gap, we first need the appropriate flat coordinate. To this end,
we solve the Picard-Fuchs equation (\ref{faeq1}) after the variable transformation 
$z\rightarrow z'=\frac{1-\D}{27}$, where $\D$ is the discriminant $\D=1-27z$. Thus, 
$\th\rightarrow\th'=(\D-1)\d_\D$ and we obtain the known closed string periods at 
the conifold. In particular, we deduce the local flat coordinate at the conifold $t_c$ 
to be,
\begin{equation}
\textstyle
t_c=\sqrt{3} \Delta +\frac{11 \Delta ^2}{6 \sqrt{3}}+\frac{109 \Delta ^3}
{81 \sqrt{3}}+\frac{9389 \Delta ^4}{8748 \sqrt{3}}+\frac{88351 \Delta ^5}
{98415\sqrt{3}}+\frac{823187 \Delta ^6}{1062882 \sqrt{3}}+\frac{68584051 \Delta^7}
{100442349 \sqrt{3}}+\cdots ~.
\end{equation}
The additional solution $\mathcal T_c$ of the inhomogeneous equation corresponds 
to the domain-wall tension at the conifold (up to a rational closed string period),
\begin{equation}
\textstyle
\mathcal T_c=\frac{\Delta ^2}{24 \sqrt{3}}+\frac{121 \Delta ^3}{2592 \sqrt{3}}+
\frac{3197 \Delta ^4}{69984 \sqrt{3}}+\frac{4372889 \Delta ^5}{100776960
   \sqrt{3}}+\frac{222720689 \Delta ^6}{5441955840 \sqrt{3}}+\frac{79384773199 
\Delta ^7}{2057059307520 \sqrt{3}}+\cdots ~.
\end{equation}
As before, we can then easily infer the expansions of $z(t_c)$, $S^{zz}(t_c)$,
and $\D^{z}(t_c)$ at the conifold point. We obtain
\begin{equation}
\begin{split}
\textstyle
z(t_c)&\textstyle =\frac{1}{27}-\frac{t_c}{27 \sqrt{3}}+\frac{11
   t_c^2}{1458}-\frac{145 t_c^3}{39366 \sqrt{3}}+\frac{6733
   t_c^4}{12754584}-\frac{120127 t_c^5}{573956280
   \sqrt{3}}+\cdots ~,\\
\textstyle
 S^{zz}(t_c)&\textstyle =-\frac{1}{1458}+\frac{4 t_c}{2187 \sqrt{3}}-\frac{103
   t_c^2}{118098}+\frac{317 t_c^3}{354294
   \sqrt{3}}-\frac{254887 t_c^4}{1033121304}+\frac{8144183
   t_c^5}{46490458680 \sqrt{3}}+\cdots ~,\\
\textstyle
\D^{z}(t_c)&\textstyle =-\frac{t_c}{324}+\frac{53 {t_c}^2}{11664 \sqrt{3}}-\frac{817{t_c}^3}{629856}
+\frac{346487 {t_c}^4}{408146688 \sqrt{3}}-\frac{17312837
   {t_c}^5}{110199605760}+\cdots ~.
\end{split}
\end{equation}
Observe that while the coordinate rescaling $t_c\rightarrow\sqrt{3} t_c$ can be used to 
make the expansions of (the closed string quantities) $z(t_c)$ and $S^{zz}(t_c)$ rational, 
the open string quantity $\D^z(t_c)$ stays irrational, therefore in comparison to the 
oriented closed string case, we do not perform such a rescaling. (Although, the rescaling
would still make the expansion of the amplitudes with an even number of boundaries
rational.) Using these expansions, we obtain the following conifold expansions of the 
amplitudes given above. 
\begin{equation}
\begin{split}
\textstyle
\F^{(2,1)}&\textstyle =-\frac{7}{466560 \sqrt{3}}+\frac{1621 {t_c}}{22394880}-\frac{97207 {t_c}^2}
{906992640 \sqrt{3}}+\frac{18202763
{t_c}^3}{587731230720}-\frac{71727601{t_c}^4}{3526387384320 \sqrt{3}}+\cdots \,,\\
\textstyle
\F^{(2,2)}&\textstyle =-\frac{227 {t_c}}{1492992 \sqrt{3}}+\frac{954653 {t_c}^2}{8707129344}-
\frac{5012287 {t_c}^3}{39182082048 \sqrt{3}}+\frac{4892098657
{t_c}^4}{135413275557888}+\cdots \,,\\
\textstyle
\F^{(2,3)}&\textstyle =\frac{545 {t_c}}{8957952}-\frac{15095299 {t_c}^2}{87071293440 \sqrt{3}}+
\frac{4878199531 {t_c}^3}{56422198149120}-\frac{92953690463
{t_c}^4}{1015599566684160 \sqrt{3}}+\cdots \,,\\
\textstyle
\F^{(2,4)}&\textstyle =-\frac{2735 {t_c}}{53747712 \sqrt{3}}+\frac{520278533 {t_c}^2}{8358844170240}
-\frac{6588078971 {t_c}^3}{56422198149120 \sqrt{3}}+\frac{1013092981
{t_c}^4}{20061226008576}+\cdots \,,
\end{split}
\end{equation}
in the oriented sector and
\begin{equation}
\begin{split}
\textstyle
\K^{(2,0)}&\textstyle =-\frac{27}{128 {t_c}^2}-\frac{47}{13824}+\frac{191 {t_c}}{279936 \sqrt{3}}+
\frac{17693 {t_c}^2}{201553920}-\frac{41893 {t_c}^3}{408146688
\sqrt{3}}+\cdots ~,\\
\textstyle
\K^{(2,1)}&\textstyle =\frac{19}{9216 \sqrt{3}}-\frac{15955 {t_c}}{35831808}-\frac{12149 
{t_c}^2}{161243136 \sqrt{3}}+\frac{29671433
{t_c}^3}{313456656384}-\frac{54115555 {t_c}^4}{626913312768 \sqrt{3}}+\cdots ~,\\
\textstyle
\K^{(2,2)}&\textstyle =-\frac{1003}{2654208}+\frac{9529 {t_c}}{17915904 \sqrt{3}}-\frac{330943 
{t_c}^2}{7739670528}-\frac{10573571 {t_c}^3}{104485552128
\sqrt{3}}+\cdots ~,\\
\textstyle
\K^{(2,3)}&\textstyle =\frac{491}{2654208 \sqrt{3}}-\frac{25373 {t_c}}{161243136}+\frac{615487 
{t_c}^2}{5159780352 \sqrt{3}}+\frac{280904809
{t_c}^3}{30091839012864}+\cdots ~,\\
\textstyle
\K^{(2,4)}&\textstyle =-\frac{193}{7077888}+\frac{191993 {t_c}}{1719926784 \sqrt{3}}-\frac{74663195 
{t_c}^2}{1486016741376}+\frac{690070327 {t_c}^3}{30091839012864
\sqrt{3}}+\cdots ~,
\end{split}
\end{equation}
in the unoriented sector. We observe that the open string amplitudes are all 
regular and $\K^{(g,0)}$ possesses similarly to $\F^{(g,0)}$ a gap at the conifold.
Namely, as $t_c\to 0$, the amplitudes are of the general form
\begin{equation}
\eqlabel{gap}
\begin{split}
\F^{(g,0)}&= \frac{\Phi_g}{t_c^{2g-2}}+ \calo(t_c^0)~, \\
{\K}^{(g,0)}&=\frac{\Psi_g}{t_c^{2g-2}}+\calo(t_c^0)~,
\end{split}
\end{equation}
the important point being that except for the leading singularity, the coefficients
of the other singular terms all vanish. Furthermore, the order of the leading singularity
at the conifold (of the amplitudes without fixed holomorphic ambiguities) can be 
easily parameterized in terms of $g$. Since we expect that this structure of the 
amplitudes is general, the holomorphic ambiguities parameterized by (\ref{FKambi}) 
need to preserve this structure. Each vanishing coefficient imposes one condition 
on $a^{(g,h)}_{\F/\K}$, \ie, fixes one coefficient $a_i$. Hence, we deduce that 
the conifold gives the following number of conditions which can be used to (partly) 
fix the ambiguities of the amplitudes:
\beq
\#_c=\left\{
\begin{matrix}
2g-3&\text{for}~\K^{(g,0)}\\
2g-2&\text{for}~\K^{(g,h)}~\text{and}~\F^{(g,1)}\\
2g-1&\text{for}~\F^{(g,h)}~(h>1)
\end{matrix}
\right.~.
\eq
Nevertheless, $\sim g$ conditions remain undetermined. In particular, the leading 
singularities of the Klein bottle amplitudes $\K^{(g,0)}$ at the conifold, which 
we have denoted as $\Psi_g$, needs to be understood. We will briefly come back 
to this point below.

One might hope that the left-over conditions can be fixed via some additional 
systematics at the orbifold point. However, performing similarly as above the expansions 
of the amplitudes at the orbifold point, we have to conclude that there is no apparent
such systematics which could aid in fixing the remaining ambiguities. 
Therefore, for the time being we have to rely on localization to fix the $\sim g$ 
remaining conditions. With the data at hand, we have completely determined $\Gx$ 
from the individual amplitudes up to $\chi=6$.  

If we instead directly compute the combined amplitude $\Gx$ via (\ref{reexpress}), we 
can go a bit further since the real topological vertex provides data for higher $\chi$. 
Similarly as for the individual amplitudes, we parameterize the holomorphic ambiguity 
of $\Gx$ via
\beq
a_{\mathcal G}^{(\chi)}=\sum_{i=0}^{n} a_i \frac{z^{i+\delta}}{(1-27z)^\zeta}~,
\eq
with $n=\frac{3}{2}\zeta$, $\delta=(\chi~\rm{mod}~2)/2$ and
\beq
\zeta=\left\{
\begin{matrix}
\chi&\text{for}~\chi~\text{even}\\
\chi-1&\text{for}~\chi~\text{odd}\\
\end{matrix}
\right.~.
\eq
The conifold expansion shows that $\Gx$ possesses a gap for $\chi$ even and is regular 
for $\chi$ odd (this is as expected from the behavior of the individual amplitudes $\F$ 
and $\K$ at the conifold described above). Similarly as for the individual amplitudes 
$\F$ and $\K$, we can easily deduce that the gap leads to 
\beq
\#_c=\left\{
\begin{matrix}
\chi-1&\text{for}~\chi~\text{even}\\
\chi &\text{for}~\chi~\text{odd}\\
\end{matrix}
\right.~,
\eq
conditions to fix the $(n+1)$ coefficients $a_i$ of $a_{\mathcal G}^{(\chi)}$ (if one 
can understand $\Psi_g$, the conifold gives exactly $\chi$ conditions).
Using the data from the real topological vertex given in table \ref{Atab3} of 
appendix \ref{secA}, we can fix the left-over conditions for some higher $\chi$ 
and in this way completely determined the amplitudes $\Gx$ up to $\chi=9$. 
\footnote{The data at hand is sufficient to go up to $\chi=12$.} The resulting 
real Gopakumar-Vafa invariants are listed in table \ref{Atab1} and \ref{Atab2} 
in appendix \ref{secA}.

Finally, let us spend a few words on the leading singularity of the $\K^{(g,0)}$ at the 
conifold \eqref{gap}. It is well known that the coefficient of the leading singularity 
of the oriented closed string amplitudes $\F^{(g,0)}$ at the conifold is given by
\cite{Ghoshal:1995wm,Ooguri:2002gx} 
\begin{equation}
\eqlabel{relationship}
\Phi_g = \frac{B_{2g}}{2g(2g-2)}~,
\end{equation}
where $B_{2g}$ are the Bernoulli numbers. The universality of the relationship 
\eqref{relationship} has been understood from many perspectives over the years. 
Among other things, $\Phi_g$ gives the Euler characteristic of the moduli space
of genus $g$ complex curves. The gap structure was discovered in 
\cite{Huang:2006si,Huang:2006hq}, and explained physically in terms of the existence 
of a single light BPS state associated with the vanishing period at the conifold
\cite{stringytest}.
It behooves us to ask for a similar interpretation of the gap structure in 
$\K^{(g,0)}$. The coefficients $\Psi_g$ have a good chance of being equally
universal as the $\Phi_g$. For future reference, we list the values of $\Psi_g$
for low $g$ in table \ref{fatab1} and leave a detailed understanding to subsequent 
work. Note that $\Psi_g$ can be conveniently extracted from ${\mathcal G}'^{(\chi)}$, 
as defined in (\ref{inteq2a}), expanded at the conifold point. This can be easily 
inferred from (\ref{inteq2}) combined with the regularity of the individual 
amplitudes with boundaries at the conifold point.

\begin{table}
\begin{center}
\begin{tabular}{|c|c|c|c|c|c|c|}
\hline
$g$&$1$&$2$&$3$&$4$&$5$&$6$\\
\hline
$\Psi_g$&$-\frac{1}{8}\log(t_c)$&$-\frac{9}{128}$&$\frac{81}{512 }$
&$-\frac{4239}{4096 }$&$\frac{221859}{16384 }$&$-\frac{48938499}{163840}$\\
\hline
\end{tabular}
\caption{$\Psi_g$ for low $g$ (note that we have rescaled $t_c\rightarrow \sqrt{3} t_c$).}
\label{fatab1}
\end{center}
\end{table}

\section{Conclusion}
\label{Conc}

In this paper, we have initiated a detailed study of the real topological string on 
local Calabi-Yau threefolds. Whereas the topological string on local (toric) Calabi-Yaus
(with toric branes) is essentially solved, and understood from a variety of different 
perspectives, and we have made significant progress on the systematics of the real
topological string, much remains to be understood (both in the local and the compact
situation). We see possibilities for further work in several directions.

The most interesting question to us is whether it is possible to achieve full integrability 
in the B-model, as is the case for the local closed topological string. As discussed 
in section \ref{fixholamb}, the behavior of the real amplitudes at the conifold point 
in moduli space does not yield enough constraints to fully fix the holomorphic
ambiguities. Therefore, it would be very desirable to find additional systematic 
constraints in order to completely fix those ambiguities. A related question is 
the interpretation of the leading singularity of the Klein bottle 
amplitudes (without boundaries) at the conifold point. One expects to be able to
find a closed expression for the leading coefficient and thereby obtain an additional 
constraint which aids in fixing the ambiguities.  

Another possible line to follow would be to generalize the real topological vertex 
presented in section \ref{realvertex} to arbitrary local toric Calabi-Yau 3-folds. 
This would put the real topological string on equal footing with the closed topological 
string (for local geometries) and would open up the arena for various case studies 
and further investigations. One might also try to generalize the recent progress
on spectral curve methods (see \cite{Eynard:2008we} for a review and references) 
as a B-model version of the topological vertex, to the real topological string.
The explicit data obtained in this work should be helpful as guideline to 
find the right formulation.

Finally, from a mathematical point of view, the localization technique originally
sketched in \cite{Walcher:2006rs,Walcher:2007qp}, and reviewed and applied in 
section \ref{oloc}, needs to be formulated in a more rigorous way (especially the 
tadpole cancellation). Also, in order to put the enumerative aspects of the 
real topological string on a firmer mathematical ground, one should seek a 
proper definition of real Gopakumar-Vafa invariants.

We believe that with the present work in hand, the real topological string can 
indeed be put on equal footing with the topological string on local geometries
in the near future. The compact case on the other hand might remain as a challenge
for some time to come. The localization and topological vertex techniques are not
applicable in the compact setting at higher genus. On the other hand, it is 
reasonable to expect that the gap structure that we found at the conifold will 
persist in compact models. This should allow for their solution to much higher
level than before. Ultimately, progress on the open sector should also feed 
back to the closed topological string. So perhaps in combination, one can learn
enough to solve both simultaneously. We look forward to further research on 
these matters.

\begin{acknowledgments}
We like to thank the organizers of the sixth Simons workshop in Mathematics and 
Physics, where this work has been initiated, for ensuring a stimulating atmosphere. 
The work of D.K. was supported in part by an EU Marie-Curie EST fellowship.
\end{acknowledgments}


\appendix

\section{\texorpdfstring{Real Gopakumar-Vafa invariants of local $\P^2$}{Real 
Gopakumar-Vafa invariants of local P2}}
\label{secA}
In this appendix, we list some real Gopakumar-Vafa invariants $N'^{(\chi)}_d$
of local $\P^2$. The results from the three complementary schemes that we
have used all agree as far as we have checked.

\begin{sidewaystable}[!hp]
\tiny
\begin{center}
\begin{tabular}{|c||c|c|c|c|c|c|}
\hline
$d$ \textbackslash ~$\chi$&$-1$&$0$&$1$&$2$&$3$&$4$\\
\hline
\hline
$1$&$1^\diamond$&&$0^*$&&$0^*$&\\
$2$&$$&$0^*$&&$0^*$&&$0^*$\\
$3$&$-1^\diamond$&&$0^*$&&$0^*$&\\
$4$&$$&$3^*$&&$1^*$&&$0^*$\\
$5$&$5^\diamond$&&$10^*$&&$6^*$&\\
$6$&$$&$-44^*$&&$-63^*$&&$-37^*$\\
$7$&$-42^\diamond$&&$-229^*$&&$-474^*$&\\
$8$&$$&$675^*$&&$2826^*$&&$6641^*$\\
$9$&$429^\diamond$&&$4833^*$&&$24547^*$&\\
$10$&$$&$-10596^*$&&$-91309^*$&&$-444825^*$\\
$11$&$-4939^\diamond$&&$-96823^*$&&$-922904^\diamond$&\\
$12$&$$&$169815^\diamond$&&$2548446^\diamond$&&$22222821^\diamond$\\
$13$&$61555$&&$1890640^\diamond$&&$29568178^\diamond$&\\
$14$&$$&$-2766312^\diamond$&&$-65141982^\diamond$&&$-907236837^\diamond$\\
$15$&$-811445$&&$-36355693$&&$-855398125$&\\
$16$&$$&$45651033$&&$1571061879$&&$32383098135$\\
$17$&$11154329$&&$692134092$&&$23061556312$&\\
$18$&$$&$-761270252$&&$-36357840387$&&$-1049953473666$\\
$19$&$-158387705$&&$-13085426739$&&$-590387680935$&\\
$20$&$$&$12804181968$&&$815896308217$&&$31671654196277$\\
$21$&$2308018713$&&$246141639751$&&$14527282829907$&\\
$22$&$$&$216905448900$&&$-17878517912137$&&$-903161239605882$\\
$23$&$-34350229129$&&$-4612322986757$&&$-346447571899667$&\\
$24$&$$&$3696709999475$&&$384413718899808$&&$24622104921447319$\\
$25$&$520291543850$&&$86171027900880$&&$8055204030496600$&\\
$26$&$$&$-63329911074864$&&$-8138918187959256$&&$-646992872220979059$\\
$27$&$-7998433661880$&&$-1606102217387496$&&$-183404890744633392$&\\
$28$&$$&$1089804320192328$&&$170128830773159693$&&$16487461934782290071$\\
$29$&$124530193132562$&&$29877825751921400$&&$4102926664405466446$&\\
$30$&$$&$-18827327577603608$&&$-3518103635914287426$&&$-409393336266808069759$\\
\hline
\end{tabular}
\caption{$N'^{(\chi)}_d$ for high $d$ obtained from the B-model (numbers marked 
with $^\diamond$ have been verified via the real topological vertex, numbers marked 
with $^*$ in addition via localization).}
\label{Atab1}
\end{center}
\end{sidewaystable}

\begin{sidewaystable}[!hp]
\tiny
\begin{center}
\begin{tabular}{|c||c|c|c|c|c|}
\hline
$d$ \textbackslash ~$\chi$&$5$&$6$&$7$&$8$&$9$\\
\hline
\hline
$1$&$0^*$&$$&$0^*$&$$&$0^*$\\
$2$&&$0^*$&$$&$0^*$&$$\\
$3$&$0^*$&$$&$0^*$&$$&$0^*$\\
$4$&&$0^*$&$$&$0^*$&$$\\
$5$&$1^*$&$$&$0^*$&$$&$0^*$\\
$6$&&$-10^*$&$$&$-1^*$&$$\\
$7$&$-497^*$&$$&$-286^*$&$$&$-91^\diamond$\\
$8$&&$9688^*$&$$&$9909^\diamond$&$$\\
$9$&$76685^*$&$$&$162007^\diamond$&$$&$240214^\diamond$\\
$10$&&$-1490889^\diamond$&$$&$-3622074^\diamond$&$$\\
$11$&$-5689826^\diamond$&$$&$-24839317^\diamond$&$$&$-80024538^\diamond$\\
$12$&&$138741207^\diamond$&$$&$660614879^\diamond$&$$\\
$13$&$309836946^\diamond$&$$&$2387676377^\diamond$&$$&$14155255239^\diamond$\\
$14$&&$-9250663299^\diamond$&$$&$-73688144692^\diamond$&$$\\
$15$&$-13813050354$&$$&$-167924131768$&$$&$-1606774464538$\\
$16$&&$496417243815$&$$&$6048297221530$&$$\\
$17$&$536811735677$&$$&$9568553947097$&$$&$136513807781008$\\
$18$&&$-22814962465032$&$$&$-399056811636330$&$$\\
$19$&$-18866208478280$&$$&$-467697511728963$&$$&$-9398297970384222$\\
$20$&&$933580323856212$&$$&$22370764847588270$&$$\\
$21$&$613983765096754$&$$&$20339969314765719$&$$&$551685003357975980$\\
$22$&&$-34902135604573377$&$$&$-1105187697763665228$&$$\\
$23$&$-18804234985799241$&$$&$-806808827756109811$&$$&$-28574033239468010587$\\
$24$&&$1213849008767132251$&$$&$49357611086785857295$&$$\\
$25$&$548264953334411255$&$$&$29708534211072505345$&$$&$1337857466210942972595$\\
$26$&&$-39792028380461566548$&$$&$-2029790874827662119329$&$$\\
$27$&$-15348471706637436099$&$$&$-1028783168774451701259$&$$&$-57640797365862616605714$\\
$28$&&$1241733288505925189151$&$$&$77934424856611454475555$&$$\\
$29$&$415237415601455194036$&$$&$33835984504174543688472$&$$&$2316194195443332049565232$\\
$30$&&$-37166974728897157340684$&$$&$-2823578149528246194259586$&$$\\
\hline
\end{tabular}
\caption{$N'^{(\chi)}_d$ for high $d$ obtained from the B-model (numbers marked with 
$^\diamond$ have been verified via the real topological vertex, numbers marked with $^*$ 
in addition via localization).}
\label{Atab2}
\end{center}
\end{sidewaystable}

\begin{table}[!hp]
\begin{center}
\tiny
\begin{tabular}{|c|ccccccccc|c|}
\hline
$\chi\setminus d$& 
 6 & 7 & 8 & 9 & 10 & 11 & 12 &13&14\\ 
\hline
10&& & 6882 & & -6527094& & 2470331689&&-472060307393\\
11  && -15 & & 254935 & & -195123249&&66336579865&\\
12  && & 3214 & & -8853482& & 7384195595&&-2473627288265\\
13 && -1 & & 195943 & & -366754317&&250379339074&\\
14  && & 988 & & -9136211& & 17862370096&&-10728530219814\\
15 && & & 109614 & & -539107092&&771890474372&\\
16  && & 191 & & -7226144& & 35296981346&&-38871359145408\\
17  & && & 44507 & & -626854392&&1965636872695&\\
18  & && 21 & & -4398773& & 57410786270&&-118572379592483\\
19  & && & 12949 & & -581661131&&4173449453891&\\
20 & && 1 & & -2061527& & 77347818109&&-306601937181157\\
21 && & & 2626 & & -433433895&&7446682383581&\\
22 & && & & -740639& & 86771638286&&-676198602671642\\
23 & && & 352 & & -260366065&&11241439498902&\\
24  & && & & -201867& & 81398541770&&-1279073229693409\\
25& && & 28 & & -126238105&&14438862544045&\\
26 & && & & -40953& & 64054115660&&-2085518321405375\\
27  & && & 1 & & -49322461&&15854057302183&\\
28  & & && & -5985& & 42371627534&&-2944249848639372\\
29 & & & && & -15453034&&14938241580054&\\
30  & & & && -595& & 23582667480&&-3613212254655871\\
31 & & & && & -3847413&&12114187918765&\\
32  & & && & -36& & 11038869636&&-3867758515991016\\
33 & & && & & -750175 &&8473209466017&\\
34  & && & & -1 & & 4337601572&&-3621885665305630\\
35 & & && & & -111971 &&5118273430606&\\
36  & & && & & & 1425576149&&-2974100596675286\\
37  & & && & & -12342 &&2671254703769&\\
38  & & && & & & 389623263&&-2145509291350998\\
39 & & & && & -946&&1204005379440&\\
40  & & & && & & 87807601&&-1361557832849019\\
41 & & && & & -45&&467997216591&\\
42 & & & && & & 16121003&&-760697816260927\\
43 & & & && & -1&&156480858834&\\
44  & & & & && & 2369885&&-374239613900020\\
45 & & & && & & &44835729183&\\
46 & & & & && & 272051&&-162059929797276\\
47  & & & & && & &10949573048&\\
48  & & & & && & 23479&&-61706256970277\\
49  & & & & && & &2262530362&\\
50  & & & & && & 1432&&-20621959046012\\
51  & & & & && & &391668488&\\
52  & & & & && & 55&&-6032986939113\\
53  & & & & && & &56047228&\\
54  & & & & && & 1&&-1539443942273\\
55 & & & & && & &6508822&\\
56  & & & & && & &&-340986604623\\
57  & & & && & & &597618&\\
58  & & & & && & &&-65152049938\\
59  & & & & && & &41728&\\
60  & & & & && & &&-10651137069\\
61 & & & & && & &2081&\\
62 & & & & & && &&-1474076916\\
63  & & & & && & &66&\\
64  & & & & && & &&-170289956\\
65 & & & & && & &1&\\
66& & & & & && &&-16111390\\
67 & & & & & && &&\\
68  & & & & & && &&-1215524\\
69  & & & & & && &&\\
70  & & & & & && &&-70301\\
71  & & & & & && &&\\
72 & & & & & && &&-2926\\
73 & & & & & && &&\\
74 & & & & & && &&-78\\
75 & & & & & && &&\\
76 & & & & & && &&-1\\
77 & & & & & && &&\\
\hline
\end{tabular}
\caption{$N'^{(\chi)}_d$ for high $\chi$ obtained via the real topological vertex.}
\label{Atab3}
\end{center}
\end{table}

\end{document}